\documentclass[aps,twocolumn]{revtex4}
\usepackage{graphicx,subfigure}
\usepackage{amsmath,amssymb}
\usepackage{color}

\newcommand{\be}{\begin{eqnarray}}
\newcommand{\ee}{\end{eqnarray}}

\newcommand{\pa}{\partial}

\newcommand{\si}{\sigma}

\newcommand{\om}{\omega}

\newcommand{\rar}{\rightarrow}

\begin{document}

\title{Fluctuations in quantum mechanics and field theories from a new version of semiclassical theory. II}

\author{M.A.~Escobar-Ruiz$^{1,2}$}
\email{mauricio.escobar@nucleares.unam.mx}

\author{E.~Shuryak$^3$}
\email{edward.shuryak@stonybrook.edu}

\author{A.V.~Turbiner$^{3,4}$}
\email{turbiner@nucleares.unam.mx, alexander.turbiner@stonybrook.edu}

\affiliation{$^1$  School of Mathematics, University of Minnesota,
Minneapolis, MN 55455, USA}

\affiliation{$^2$  Centre de Recherches Math\'ematiques, Universit\'e de Montreal,
C.P. 6128, succ. Centre-Ville, Montr\'eal, QC H3C 3J7, Canada}

\affiliation{$^3$  Department of Physics and Astronomy, Stony Brook University,
Stony Brook, NY 11794-3800, USA}

\affiliation{$^4$ Instituto de Ciencias Nucleares, Universidad Nacional Aut\'onoma de M\'exico,
Apartado Postal 70-543, 04510 M\'exico, D.F., M\'exico}

\date{\today}

\begin{abstract}
This is the second paper on semiclassical approach based on the density matrix given by the Euclidean time path integral with fixed coinciding endpoints. The classical path, interpolating between this point and the classical vacuum, called ``flucton", plus systematic one- and two-loop corrections, has been calculated in the first paper \cite{Escobar-Ruiz:2016aqv} for double-well potential and now extended for a number of quantum-mechanical problems (anharmonic oscillator, sine-Gordon potential).
The method is based on systematic expansion in Feynman diagrams and thus can be extended to QFTs.
We show that the loop expansion in QM reminds the leading log-approximations in QFT.
In this sequel we present complete set of results obtained using this method in unified way.
Alternatively, starting from the Schr\"{o}dinger equation we derive a {\it generalized} Bloch equation which semiclassical-like, iterative solution generates the loop expansion. We re-derive two loop expansions for all three above potentials and now extend it to three loops, which has not yet been done via Feynman diagrams. All results for both methods are fully consistent with each other. Asymmetric (tilted) double-well potential (non-degenerate minima) is also studied using the second method.
\end{abstract}

\maketitle

\section{Introduction }
Semiclassical approximations are well known tools, both in quantum mechanical and quantum field theory. Standard textbooks of quantum mechanics usually start with Bohr-Sommerfeld quantization conditions, and
semiclassical WKB approximation  for the wave function. Unfortunately, extending such methods beyond of the first correction for the one-dimensional case or those with separable variables, or to a multidimensional case, proved to be difficult.

In our previous paper \cite{Escobar-Ruiz:2016aqv}, to be referred below simply as I,
we introduced a different version of the semiclassical approach, based on
Feynman's path integral representation of the density matrix
\cite{FH_65,Feynman_SM} analytically continued to imaginary (Euclidean) time.
It corresponds to a transition from quantum mechanics to statistical mechanics:
If the time is defined as a periodic variable, with period $\beta=\hbar/T$, the density matrix
corresponds to quantum ensemble at nonzero temperature $T$. However in this paper
we will  only focus on  the zero temperature limit, in which the period is infinite.


At the classical level, the theory is based on a classical (minimal action) periodic path,
which extends from some arbitrary point $x_0$ to the ``classical vacuum", the minimum of the potential, and return. This path has been introduced in \cite{Shuryak:1987tr} and was named ``flucton".

General advantage of this approach is that the path integrals lead to systematic perturbative series, in the form of Feynman diagrams, with clear rules for each order. Text-book perturbative approaches for the wave functions do not have that, and basically are never used beyond say first and second orders.

Of course,  the higher level of generality comes with a heavy price. While classical part is relatively simple, already quantum part at one-loop level one needs to calculate determinants of certain differential operators. At two and more loops Feynman diagrams need to be evaluated on top of  space-time dependent backgrounds: therefore those should be evaluated in the space-time representation rather than in energy-momentum one mostly used in QFT applications. Most content of the first part of this paper is the explicit demonstration of how one can do all that, in analytic form, for three classical examples -- quartic anharmonic and double-well, and sine-Gordon (Mathieu) potentials. Their quantum Hamiltonian is of the standard form
\be
\label{Hamiltonian}
{\cal H}\ =\ -\frac{1}{2m} \pa_x^2 \ +\ V(x)\ ,\ \pa_x=\frac{d}{dx} \ ,
\ee
where we use units $\hbar=1$ and $m=1$, without a loss of generality.

Since our ultimate aim remains a generalization of the semiclassical theory to QFT's, our quantum-mechanical examples should be represented in a certain specific form of anharmonic perturbation of the harmonic oscillator
\be
\label{potential}
                 V(x)\ =\ \frac{\tilde V(g x)}{g^2}\ =\ \frac{1}{2}\, x^2 + a_3\, g x^3 + a_4\, g^2 x^4 + \ldots
\ee
where $\tilde V$ has a minimum at $x=0$, it always starts from quadratic terms, the frequency of the small oscillation near minimum placed equal to one, $\om=1$ and $g$ is the coupling constant. Classical (vacuum) energy is always taken to be zero, $V(0)=0$ and $a_{2,3,\ldots}$ are parameters. We call $(g x)$ the {\it classical} coordinate, see below. Both the classical coordinate and the Hamiltonian (\ref{Hamiltonian})
are invariant with respect to simultaneous change
\[
     x \rar -x \quad , \quad g \rar -g\ .
\]
It implies that the energy is the function of $g^2$,
\be
\label{energy}
    E\ =\ E(g^2)\ .
\ee
The semiclassical expansion is done in powers of small coupling $g$. In a way it is similar to the perturbation theory which is also in powers of small coupling $g$.

Let us indicate potentials we are going to study in the form (\ref{potential}):

Quartic AnHarmonic Oscillator (AHO),
\be
\label{AHO}
                 V\ =\ \frac{1}{2}\,x^2 (1 + g^2 x^4)\ ,
\ee

Quartic Double Well Potential (DWP),
\be
\label{DWP}
                 V\ =\ \frac{1}{2}\,x^2 (1 - g x)^2\ ,
\ee

Sine-Gordon (Mathieu) Potential (SGP),
\be
\label{SGP}
                 V\ =\ \frac{(1 - \cos {g x})}{g^2}\ ,
\ee
and eventually,

Quartic Asymmetric (tilted) Double Well Potential (ADWP):
\be
\label{ADWP}
                 V\ =\ \frac{1}{2} x^2 (1 + 2t g x + g^2 x^2)\ ,
\ee
where the parameter $t$ ``measures" the asymmetry of wells, for $t=\pm 1$ the wells are symmetric and we arrive at DWP, while for $t=0$ there occurs AHO.

In our previous paper \cite{Escobar-Ruiz:2016aqv} we used Feynman diagrams to calculate one and two-loop corrections to classical flucton action for DWP potential. For two other famous quantum-mechanical problems, the AHO and SGP, we had only presented the derivation of the Green functions and the (one-loop) determinants. For completeness, here we also add the results for the two-loop corrections, calculated from the same set of Feynman diagrams, for those two problems. Like for DWP case the complete set of results looks surprisingly simple and compact. Remarkably, it does not contain any transcendental functions, logs and polylogs, which appear for individual diagrams.
Furthermore, the classical (flucton) action is always the WKB action, $\int p dq$, but
evaluated at $zero$ energy - this general observation was missed before.
To clarify a meaning of loop expansion we will be able to derive a certain Bloch-type, Riccati equation the iteration solution of which generates exactly the loop expansion. This equation will be called {\it the generalized Bloch equation}.

\section{Fluctuation corrections from the Feynman diagrams}

The method has been extensively described in I and there is no need to repeat it here in detail.
Its main idea is that quantum fluctuations around classical flucton path $x_{\text{fluction}}$ can be
described by standard expansion of the action in the powers of $(x-x_{\text{fluction}})$, with quadratic
term giving by the Green function while the higher order terms produce vertices of the Feynman diagrams.
Let us only remind in brief the main definitions.

\subsection{Generality}
\label{sec_setting}
By definition the Feynman path integral gives the density matrix in quantum mechanics
\cite{FH_65}
\be
\label{denmatr}
\rho(x_i,x_f,t_{tot})\ =\ N\,\int_{x(0)=x_i}^{x(t_{tot})=x_f} Dx(t) e^{i\,S[x(t)]/\hbar} \ .
\ee
Here $N$ is a normalization factor and $S$ is the usual classical action of the problem,
\[
S \ = \ \int_0^{t_{tot}}dt\, \bigg[ \frac{m}{2}{\bigg(\frac{dx}{dt}\bigg)}^2 - V(x)    \bigg] \ ,
\]
for a particle of mass $m$ in a static potential $V(x)$ - it provides the weight of the paths in (\ref{denmatr}).

As it is well known, see e.g.\cite{Feynman_SM}, one can also apply these expressions
in statistical mechanics. For this one needs to change time into its Euclidean version $\tau=i\,t$ defined  on a circle with circumference $\beta=\tau_{tot}$. Such periodic time is known as the Matsubara time, and the density matrix of quantum system is related to probability for thermal system with temperature
\be
\label{time}
      T\ =\ \hbar/\beta \ .
\ee
At $T \rar 0$  the ground state of the quantum system is naturally recovered.
Periodicity of the path implies that there is only one endpoint $x_i=x_f=x_0$.

The main object of our study is the diagonal matrix element of the density matrix, giving the probability
for the specific coordinate value $x_0$ (or a particular field configuration $\phi_0(\vec x)$ in QFT) in this ensemble
\be
P(x_0,\beta) \ =\ N\,\int_{x(0)=x_0}^{x(\beta)=x_0} Dx(\tau) e^{-S_E[x(\tau)]/\hbar} \ .
\label{Px}
\ee
So, we take into account all (closed) trajectories starting and ending at $x_0$.
Here the weight is defined via the Euclidean action
\[
   S_E \ = \ \int_0^{\beta}d\tau\, \bigg[ \frac{m}{2}{\left(\frac{dx}{d\tau}\right)}^2 + V(x)\bigg] \ .
\]
Using standard definition of the density matrix in terms of stationary states $\mid n \rangle$ with energy $E_n$, the sum over states becomes a set of decreasing exponentials
\be
\label{P0}
P(x_0,\beta)=\sum_n |\psi_n(x_0)|^2 e^{-E_n \beta} \ .
\ee
In the limit of large $\beta$ or low temperature $T$, in the expression (\ref{P0}) the dominant term
\be
P(x_0,\beta \rightarrow \infty) \sim |\psi_0(x_0)|^2\ e^{-E_0 \beta} \ ,
\label{psi0}
\ee
describes the ground state, the main state we are interested in.

\subsection{The classical path - flucton}
\label{sec_flucton}

We assume for simplicity that the potential (\ref{potential}) has a global minimum at $x=0$,
\[
V(x) \ \geq \ 0  \qquad \text{and} \qquad \frac{d}{d\,x}V|_{x=0} \ = \ 0 \ .
\]
thus, the exponent in (\ref{P0}) is non-negative, $S_E \geq 0$.

In Euclidean time $\tau$ the kinetic energy changes sign, which is equivalent to
the potential effectively  flipping  sign, $V(x)\rightarrow -V(x)$, turning a
minimum into a maximum. Now, let us ask if there exist a real path, starting at some
arbitrary point $x_0$ at $\tau=0$ and
returning to it after the required time duration, at the Matsubara time $\tau=\beta$.
The lowest action path of this kind is the classical path we call $flucton$.
Its energy is defined by its period.

\begin{figure}[h!]
\begin{center}
\includegraphics[width=3.0in,angle=0]{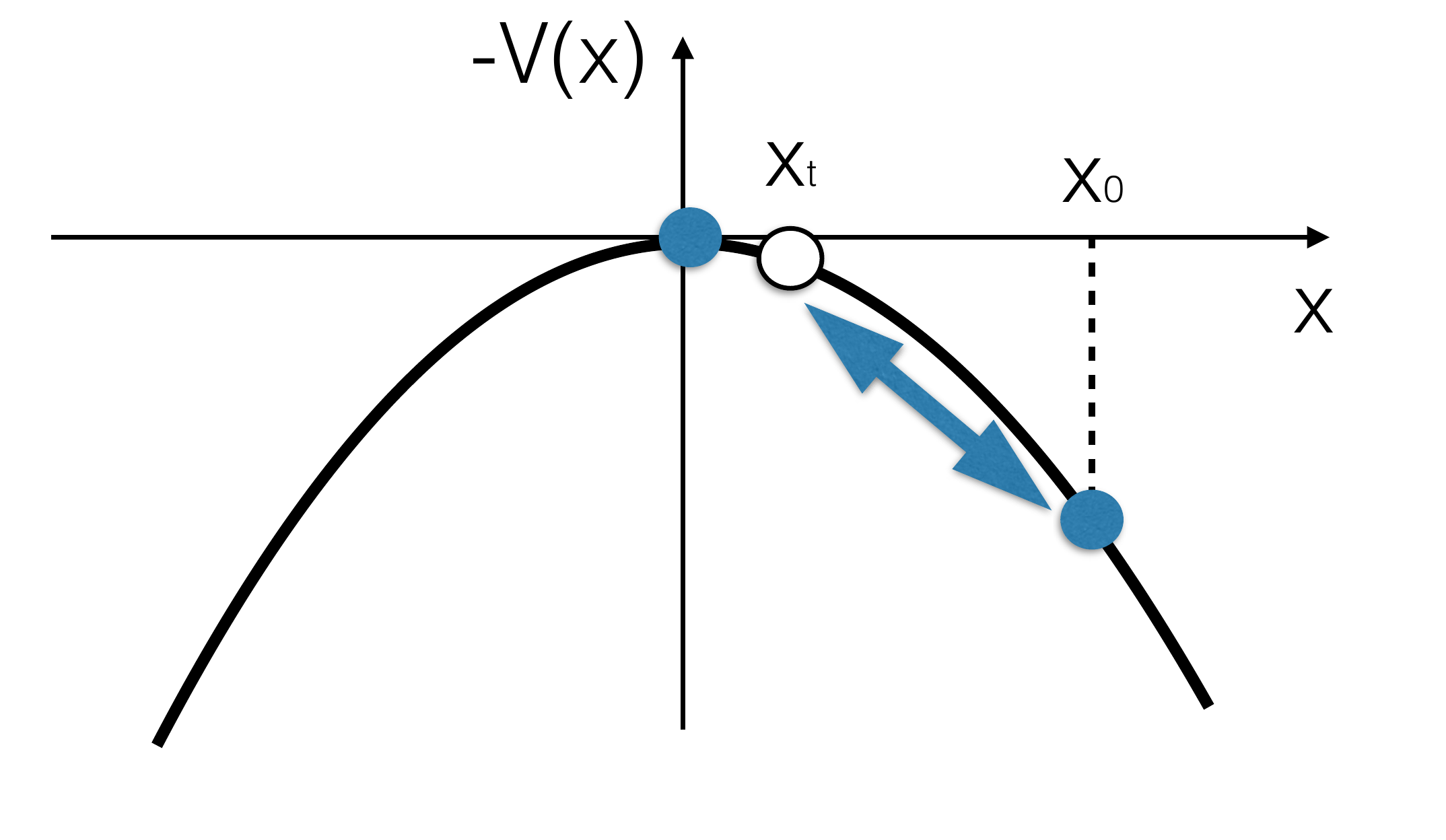}
\caption{The sketch of the inverted potential $-V$ versus coordinate $x$.
Flucton is the classical trajectory starting and ending at the same  initial point $x_0$. At non-zero temperature it goes through the turning point $x_t$, see text. At zero temperature $x_t$ coincides with the location of the maximum, $x_t=0$.}
\label{flucton}
\end{center}
\end{figure}

Since in this work we deal only with quantum mechanical limit of vanishing temperature  $T\rightarrow 0$,
the Matsubara time goes to infinity.
It is clear then that the particle should spend a divergently long time near the
turning point, which is the case when $x_t\rightarrow 0$, the location of the maximum of $-V$, see Fig.\ref{flucton}.
Evidently, such classical path with infinite period is the one
with zero energy $E=0$.  The basic idea is that  such classical path with zero energy $E=0$ "climbs up the hill" to its maximum at $x=0$.

Let us find the flucton paths explicitly.
They, of course, satisfy the second order classical Equation Of Motion (EOM),
but in the one-dimensional case
it is much easier to use the energy conservation, at $E=0$
\be
\label{EOM-N}
\frac{d}{d\tau}x(\tau) \ =  \ \sqrt{2\,V(x)} \ .
\ee
The circular trajectory emerging in (\ref{EOM-N}) which starts and ends at $x_0$ passing through $x=0$ for the time $\beta$ is called a {\it flucton} \cite{Shuryak:1987tr},
\[
x_{\text{flucton}}(0)=x_{\text{flucton}}(\beta)=x_0 \ ,
\]

\[
x_{\text{flucton}}\ =\ x_{\text{flucton}}(\tau;\,x_0,\,\beta) \ .
\]
It enables us to evaluate the transition amplitude $P(x_0,\beta)$ (\ref{Px}). Putting
\[
\phi(x_0\,,\beta) \ \equiv \ -\log[P(x_0,\beta)] \ ,
\]
in $(\ref{psi0})$ and expanding the classical action around the flucton we obtain
\[
  \phi(x_0\,,\beta) \ =
\]
\be
\label{Pfluc}
   S_{\text{flucton}} + \frac{1}{2}\log (N^{-2}\,\text{det}(O_{\text{flucton}}))\ +\ \text{loops} \ ,
\ee
where $S_{\text{flucton}} \equiv S_E[x_{\text{flucton}}]$ and

\be
\label{Oflucton}
  O_{\text{flucton}} \ \equiv \ -\frac{d^2}{d^2\tau} + {\frac{\pa^2}{{\pa x}^2 } V(x)}|_{x=x_{\text{flucton}}(\tau;\,x_0,\,\beta)}  \ ,
\ee
is a Schr\"{o}dinger-type operator in $\tau$ variable with $x_0, \beta$ as parameters.

In order to construct the loop expansion (\ref{Pfluc}) three building blocks are used: \\
(i) The action $S_{\text{flucton}}$ of the flucton. In the limit $\beta \rightarrow \infty$, the action $S_{\text{flucton}}$ provides the dominant term of the phase of the ground state function (see (\ref{psi0})) and reproduces exactly the WKB result (obtained from the Riccati-Bloch equation for the phase).\\
(ii) The determinant of $O_{\text{flucton}}$ (\ref{Oflucton}) describes the quadratic quantum fluctuations.\\
(iii) The loop-corrections, true quantum corrections decreasing at large distances and in the present formalism given by explicit Feynman diagrams in the flucton-background.\\
While the computation of $S_{\text{flucton}}$ (\ref{Pfluc}) is relatively simple, the evaluation of $\text{det}O_{\text{flucton}}$ already involves the diagonalization of a certain non-trivial second-order differential operator. Usually, it is a highly non-trivial calculation which is enormously simplified by the generalized Riccati-Bloch equation, see below.

While in our paper I we discussed to some extent several quantum mechanical problems,
only for the double well potential (DWP) the calculations have been done to two loops. It is clear now
that the formalism can be applied for {\it any} potential of a type of perturbed harmonic oscillator (\ref{potential}). Therefore, without explanations, we now list simply the full set of the results,
for AHO, DWP and SGP potentials and also for ADWP.
The units used assume particle mass $m=1$ and the Planck constant $\hbar=1$.

\subsubsection{\large Relating the determinant and the Green function
\label{sec_det_GF} }

In this section we calculate the quadratic order quantum oscillations around classical (flucton) path, namely, the determinant (\ref{Pfluc}).
For the harmonic oscillator, the potential  ${\pa^2 V(x)/{\pa x}^2 }|_{x=x_{\text{flucton}}}$ in (\ref{Oflucton}) is just a constant, so in this case the fluctuations do not depend on the classical path, and direct diagonalization of the operator (\ref{Oflucton}) \cite{Vainshtein:1981wh} shows that
\[
N^{-2}\,\text{Det}(O_{\text{flucton}})\ = \ 2\,\pi\,\sinh \beta   \ .
\]
In general, a direct diagonalization of (\ref{Oflucton}) is highly-non trivial and analytical results
for it are extremely rare.

\begin{figure}[h]
\begin{center}
\includegraphics[width=3.0in,angle=0]{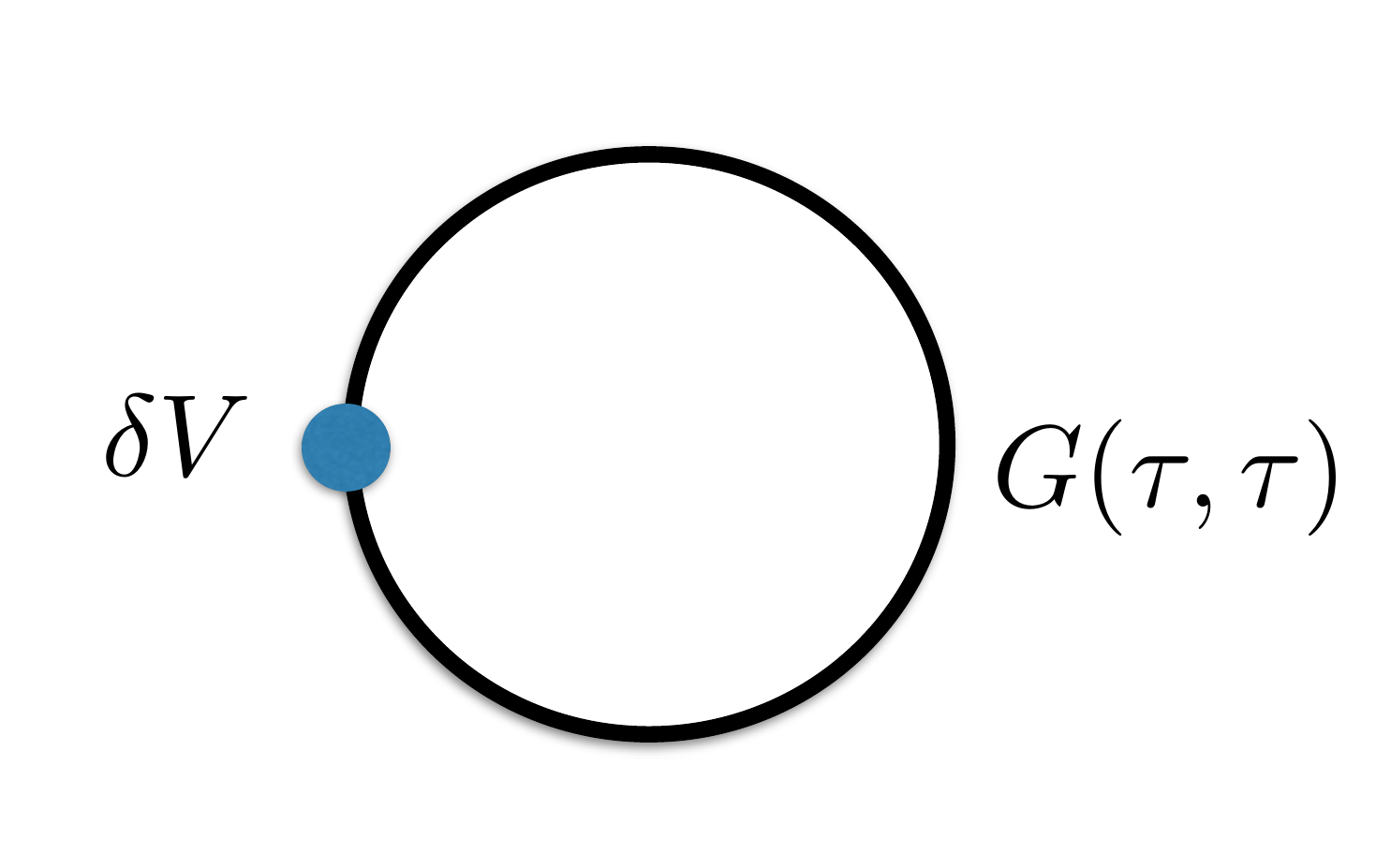}
\caption{Symbolic one-loop diagram, including variation of the fluctuation potential $\delta V$ and the simplified ``single-loop" Green function $G(\tau,\tau)$\,, see \cite{Escobar-Ruiz:2016aqv}.
}
\label{fig_oneloop}
\end{center}
\end{figure}

At 1978 Brown and Creamer \cite{Brown:1978yj} invented the way how to relate the determinant and the Green function reducing it to calculation of a symbolic one-loop Feynman diagram. One can apply their procedure
to quantum mechanics.
When the potential $V_{\text{flucton}}\equiv V(x_{\text{flucton}}(\tau;\,x_0,\,\beta))$ depends on some parameter, it can be varied. To this end, we rewrite the potential as
\[
    V_{flucton}\ =\ 1\ +\ W(\tau;\,X,\,\beta) \ ,
\]
where $X=X(x_0)$. Its variation resulting in an extra potential
\be
\label{deltaV}
     \delta  V_{\text{flucton}}\ =\ \frac{\pa W}{\pa X} \delta X
\ee
is a perturbation: its effect can be evaluated by the following Feynman diagram
\be
\label{relation}
 \frac{\pa \log \text{Det}\,(O_{\text{flucton}})}{\pa X}\ =\ \int d\tau G(\tau,\tau)
 \frac{\pa V_{\text{flucton}}(\tau)}{\pa X}\ ,
\ee
containing derivative of the potential as a vertex and the ``loop''-Green function $G(\tau,\,\tau)$\,,
\be
\label{GFfluc}
  O_{\text{flucton}}\,G(\tau_1,\tau_2)  \ = \ \delta(\tau_1 - \tau_2)  \ ,
\ee
at the same point $\tau_1=\tau_2=\tau$, see Fig.\ref{fig_oneloop}. For simplicity, the dependence of $G(\tau_1,\,\tau_2)$ on $x_0$ and $\beta$ is omitted. The equation (\ref{relation}) relates the determinant and the Green function: if the r.h.s. of it can be calculated, the derivative over $X$ can be integrated back.
Hence, if the Green function is known, one can calculate the determinant.

\subsubsection{\large Two-loop-correction: Feynman diagrams}

\label{sec_higher_loops}

The loop expansion for $P(x_0, \beta)$ can be written
in the form
\be
\text{loops} \ = \ 1 + g^2\,B_1 + g^4\,B_2+...
\label{loopcorr}
\ee
where $B_n=B_n(x_0)$ is the $n$-loop contribution. Equivalently,
in (\ref{Pfluc}) the loop expansion is of the form
\be
\text{loops} \ = \ g^2\,B_1 + g^4\,{\tilde B}_2 + ...
\label{loopcorr2}
\ee
where ${\tilde B}_n={\tilde B}_n(x_0)\ ,\ n>1$ is made out of loop contributions $B_n$, e.g.
${\tilde B}_2 = -B_1^2/2 + B_2$\ etc.

\begin{figure}[h!]
\begin{center}
\includegraphics[width=3.0in,angle=0]{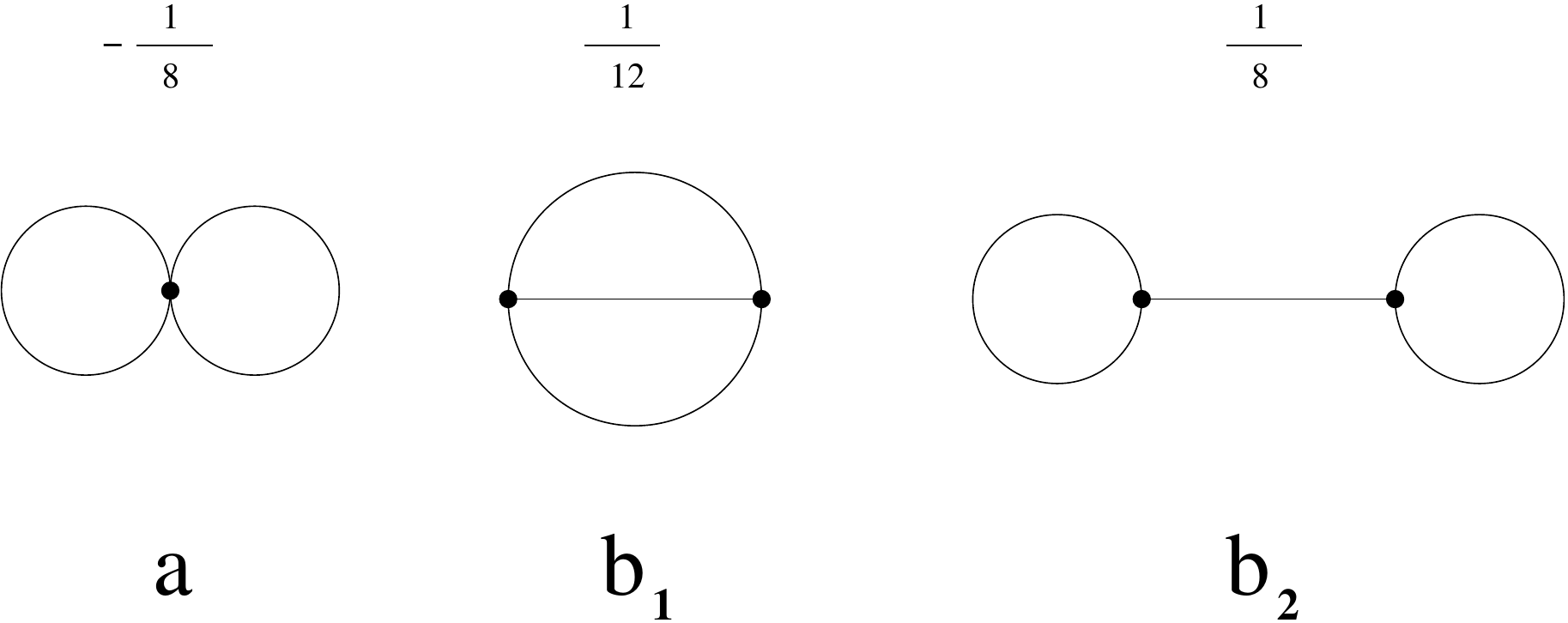}
\caption{Diagrams contributing to the two-loop correction $B_1 = a + b_1 + b_2$. The signs of contributions and symmetry factors are indicated. }
\label{2loopplot}
\end{center}
\end{figure}

In general, for any potential of the form (\ref{potential}) the two-loop correction $B_1$ is given by the sum of three Feynman diagrams, see Fig. \ref{2loopplot}\,,

where diagram $a$ is given by a one-dimensional integral while the diagrams $b_1$ and $b_2$ correspond to two-dimensional integrals. Explicitly,
\be
 a \ \equiv \ -\frac{1}{8}\,\int_0^{\infty}[v_4(\tau)\,G^2(\tau,\,\tau) -
 v_{4,0}\,G_0^2(\tau,\,\tau) ]d\tau \ ,
\nonumber
\ee

\be
 b_1 \ \equiv \ \frac{1}{12}\,\int_0^{\infty}\int_0^{\infty}[v_3(\tau_1)\,v_3\,(\tau_2)G^3(\tau_1,\,\tau_2)
\nonumber
\ee

\[
 -\ v_{3,0}v_{3,0}G_0^3(\tau_1,\,\tau_2) ]\,d\tau_1\,d\tau_2 \ ,
\]

\be
b_2 \ \equiv \ \frac{1}{8}\,\int_0^{\infty}\int_0^{\infty}\big[v_3(\tau_1)\,v_3\,
     (\tau_2)G(\tau_1,\,\tau_1)G(\tau_1,\,\tau_2)G(\tau_2,\,\tau_2)
\nonumber
\ee

\be
-\ v_{3,0}v_{3,0}G_0(\tau_1,\,\tau_1)G_0(\tau_1,\,\tau_2)G_0(\tau_2,\,\tau_2)\big]
   \,d\tau_1\,d\tau_2 \ ,
\label{diagrams}
\ee
where

\[
v_{k}(\tau) \ = \ {\frac{\pa^k}{{\pa x}^k } V(x)}|_{x=x_{\text{flucton}}} \ , \qquad k=3,4
\]
are the vertices in the flucton background,

\[
v_{k,0} \ = \ {\frac{\pa^k}{{\pa x}^k } V(x)}|_{x=0} \ , \qquad k=3,4
\]
denote the ``vacuum vertices'', the Green function $G(\tau_1,\,\tau_2)$ is defined in (\ref{GFfluc}) and

\be
\label{GDW0}
G_0\ =\ \frac{e^{-|\tau_1 - \tau_2|}}{2} - \frac{e^{-\tau_1 - \tau_2}}{2} \ ,
\ee
is the ``harmonic propagator". Its presence is related with the necessity to subtract
(space-time divergent) anharmonic effects, unrelated to the fluctons.

In the case of three-loop contribution for general potential (\ref{potential})
there exist 15 diagrams which contribute, see e.g. \cite{Escobar-Ruiz:2015rfa}, while for DWP the number of
diagrams drops to 12, see e.g. \cite{Escobar-Ruiz:2015nsa}. Like in instanton calculus we do not hope all these diagrams can be calculated analytically.

\subsection{AHO}

This section contains the expansion (\ref{Pfluc}) up to two-loops for AHO.

The {\em anharmonic oscillator} (AHO) is defined by the potential (\ref{AHO})
\[
  V(x) \ = \ \frac{1}{2}  x^{2}\ (1+g^2\,x^2) = \frac{1}{2g^2}  u^{2}\ (1+u^2)\ ,\ u=g x\ ,
\]
at zero temperature, $T=0$. The classical flucton path solution with the energy $E=0$ of (\ref{EOM-N}) is

\[
x_{\text{flucton}}(\tau) \ =\ \frac{g\,x_0}{ \cosh(|\tau|) +\sqrt{1+g^2\,x_0^2}\,\sinh(|\tau|)} \ .
\]
The corresponding classical action is given by

\[
S_E[x_{\text{flucton}}]\ =\ \frac{2}{3}\,\frac{{(1+g^2\,x_0^2)}^{\frac{3}{2}} -1}{g^2} \ .
\]
In the limit $g \rightarrow 0$ we recover the classical action for the harmonic oscillator and at $x_0\rightarrow \infty$ we obtain the expansion
\be
  S_E[x_{\text{flucton}}] \ = \   \frac{2}{3}\,g\,x_0^3+\frac{1}{g}x_0\,+\,\text{lower order terms}
\ee

It is convenient to introduce a new variable
\be
\label{XAHO}
  X_{\small AHO}(x)\ =\ {(1+g^2\,x^2)}^{\frac{1}{2}}\ =\ {(1+u^2)}^{\frac{1}{2}}\ ,\ u=g x\ .
\ee

For the anharmonic oscillator, the Green function of the operator $O_{\text{flucton}}$ (\ref{Oflucton}) is given by

\begin{widetext}
\begin{equation}
\begin{aligned}
G(\tau_1,\,\tau_2;X_{\small AHO}) & \ = \
     \frac{\mbox{sech} [\frac{1}{2} (-|\tau_-| + \tau_+)] \left(X_{\small AHO} \cosh [\frac{1}{2} (|\tau _-|+\tau _+)]+\sinh [\frac{1}{2} (|\tau_-| + \tau_+)]\right)}
          {4X_{\small AHO} \left(\cosh [\frac{1}{2} (|\tau_-| + \tau_+)] + X_{\small AHO} \sinh [\frac{1}{2} (|\tau_-|+\tau_+)]\right){}^2 \left(1+X_{\small AHO} \tanh [\frac{1}{2} (-|\tau_-|+\tau_+)]\right){}^2}
\\
& \times
   \bigg[ X_{\small AHO} \cosh [|\tau_-| - \tau_ +] \left(1+3 X_{\small AHO}^2+X_{\small AHO} (3+X_{\small AHO}^2 ) \tanh [\frac{1}{2} (-|\tau_-|+\tau_+)]\right)
\\
& + (1-X_{\small AHO}^2) \left(4-5 X_{\small AHO}^2-3 X_{\small AHO} (|\tau_-| - \tau_+)\right)
  \tanh [\frac{1}{2} (-|\tau_-|+\tau_+)]
\\
&  -\ X_{\small AHO}
    \bigg(1+3 X_{\small AHO} \left(X_{\small AHO}+(1-X_{\small AHO}^2) (| \tau_- | - \tau_+ )\right)\bigg)\bigg]
\end{aligned}
\end{equation}
where $\tau_-=\tau_2-\tau_1\ ,\ \tau_+=\tau_1+\tau_2$\,.

Taking the above Green function and the ``vertex"
\be
   \frac{\pa V_{\text{flucton}}(\tau)}{\pa X_{\small AHO}}\ =\
   \frac{12 \left(\sinh (\tau)\,+\, \,X_{\small AHO}\,\cosh (\tau)\right)}{\left(\cosh (\tau)\, + \, X_{\small AHO}\,\sinh (\tau)\right){}^3} \ ,
\ee
$(\tau>0)$ to evaluate (\ref{relation}) we obtain analytically

\be
\label{detAHO}
      \log \, \text{Det} (O_{\text{flucton}})\ =\ 2 \log[X_{\small AHO} (1+X_{\small AHO})] \ .
\ee

As for the next term, the two-loop correction $B_1$, the results for all three two-loop Feynman diagrams shown in Fig.\ref{2loopplot}, can be found analytically and they are

\begin{equation}
\label{aAHO}
\begin{aligned}
& a\ =\  -\frac{3 (32+96 X + 29 X^2-74 X^3-35 X^4)}{560 X^2 (1+X)^2} \\
& \qquad -\ \frac{9\bigg( 2(2-3 X^2) \log(2) - 2(2-3 X^2) \log[1+X]-3 X (1-X^2) \text{PolyLog}[2,\frac{X-1}{1+X}]\bigg)}{70 X (1-X^2)} \ , \\ \\
&    b_1 \ =\ -\frac{140+184 X+272 X^2+193 X^3-478 X^4-455 X^5}{840 X^3 (1+X)^2} \\
& \qquad   +\frac{3 (2(2-3 X^2) \log(2) - 2 (2-3 X^2) \log [1+X]-3 X (1-X^2) \text{PolyLog}[2,\frac{X-1}{1+X}])}{35 X (1-X^2)} \ , \\  \\
&   b_2 \ =\ -\frac{140+528 X+464 X^2-169 X^3-626 X^4-385 X^5}{560 X^3 (1+X){}^2} \\
& \qquad +\frac{3 (2(2-3 X^2) \log(2) - 2 (2-3 X^2) \log [1+X]-3 X (1-X^2) \text{PolyLog}[2,\frac{X-1}{1+X}])}{70 X (1-X^2)} \ ,
\end{aligned}
\end{equation}
\end{widetext}
where $X\equiv X_{\small AHO}$ and $\text{PolyLog}[n,z] =\sum_{k=1}^{\infty} z^k/k^n$ is the polylogarithm. Each diagram provides a contribution in a form of sum of rational (meromorphic) function and a transcendental function. Functionally, both functions are similar in each diagram.
Eventually, after summing all three diagrams $a, b_1, b_2$, the two-loop correction $B_1$ takes an amazingly simple form,

\begin{widetext}
\be
\label{B1AHO}
  B_1^{(AHO)}\ = \ \frac{(1-X_{\small AHO})(5+16 X_{\small AHO}+25 X_{\small AHO}^2+17 X_{\small AHO}^3)}{12 X_{\small AHO}^3 (1+X_{\small AHO})}\ ,
\ee
\end{widetext}
where all transcendental contributions are {\em cancelled out(!)} and the answer turns out to be the meromorphic function of $X_{\small AHO}$ {\it only}. We will observe similar cancellations for DWP and SGP below.

\subsection{DWP}

The {\em double well potential} we define as in (\ref{DWP})

\[
 V \ = \  \frac{1}{2}x^2\,( 1 - g\,x\,)^2 \ =\ \frac{1}{2g^2}  u^{2}\,(1-u)^2\ ,\ u = g\,x\ .
\]
Its two degenerate minima are situated at $x=0$ and $x= 1/g$, respectively. At zero temperature
limit $T=1/\beta=0$, the flucton trajectory is given by \cite{Escobar-Ruiz:2016aqv}

\[
x_{\text{flucton}}(\tau) \ =\ \frac{x_0\,(\cosh(|\tau|)-\sinh(|\tau|)) }{1+g \,x_0\,(1-\cosh(|\tau|)+\sinh(|\tau|))}  \ ,
\]
and the corresponding classical action reads
\[
S_E[x_{\text{flucton}}]  \ =\  x_0^2\, (1 + \frac{2 \,g\, x_0}{3}) \ .
\]

For this case it is convenient to introduce the variable
\be
\label{XDWP}
  X_{\small DWP}(x)\ \equiv\ u\ =\ g\,x\ ,
\ee
c.f. (\ref{XAHO}).
In this variable the corresponding Green function takes the form
\begin{widetext}

\[
    G(\tau_1,\,\tau_2)\ =\ \frac{e^{-|\tau_1-\tau_2|}}{2\ {\big(e^{\tau_1}(1+X)-X\big)}^2\,{\big(e^{\tau_2}(1+X)-X\big)}^2}
    \bigg[ 8\ e^{\frac{1}{2}(\tau_1+\tau_2+3\,|\tau_1-\tau_2|)}\,X^3\,(1+X)
\]
\[
    -\ 8\ e^{\frac{1}{2}(3\tau_1+3\tau_2+|\tau_1-\tau_2|)}\,X\,(1+X)^3
    +\ e^{2\,(\tau_1+\tau_2)}\,{(1+X)}^4      -6\,e^{(\tau_1+\tau_2+|\tau_1-\tau_2|)}\,X^2\,{(1+X)}^2\, |\tau_1-\tau_2|
\]
\[
   +\  e^{(\tau_1+\tau_2+|\tau_1-\tau_2|)}\,\bigg(\,6\,X^4\,(\tau_1+\tau_2)\, +\, 12\,X^3\,(1+\tau_1+\tau_2)
 +\ 6\,X^2\,(3 + \tau_1+\tau_2)  + 4\,X - 1 \bigg)
\]
\be
\label{GDW}
  - e^{2\,|\tau_1-\tau_2|}\,X^4 \ \bigg] \ ,
\ee
\end{widetext}
for $\tau_1,\,\tau_2>0$\,, where $X=X_{\small DWP}$.

Substituting (\ref{GDW}) and the ``vertex"
\be
  \frac{\pa V_{\text{flucton}}(\tau)}{\pa X_{\small DWP}}\ =\
    \frac{6 e^\tau \big(X_{\small DWP} + e^\tau (1 + X_{\small DWP})\big)}{\big( e^\tau (1 + X_{\small DWP}) - X_{\small DWP}\big)^3}\ .
\ee
in the r.h.s. (\ref{relation}) gives

\be
\label{detDWP}
     \frac{\pa \log \text{Det}\,({O_{\text{flucton}} })}{\pa X_{\small DWP}}\ =\
     \frac{4}{1 + X_{\small DWP}}\ ,
\ee
c.f. (\ref{detAHO}).

The results for all three two-loop Feynman diagrams shown in Fig.\ref{2loopplot} can be found analytically

\begin{widetext}
\[
 a \   =\ \frac{3}{560 X^2 (1+X)^4}  \times \bigg(24X-60X^2-520X^3 -1024 X^4 - 832 X^5 - 245 X^6
\]
\[
 -24 (1+X)^2 (1-4 X-18 X^2-12 X^3) \log[1+X] +288 X^2 (1+X)^4\, \text{PolyLog}[2,\frac{X}{1+X}]\bigg) \ ,
\]

\be
\label{aDWP}
 b_1 \ =\  -\frac{1}{280 X^2 (1+X)^4} \times
  \bigg(24X-60 X^2-520 X^3-1024X^4-832 X^5-245 X^6
\ee
\[
 - 24 (1+X)^2 (1-4 X-18 X^2-12 X^3) \log[1+X] + 288 X^2 (1+X)^4\, \text{PolyLog}[2,\frac{X}{1+X}]\bigg) \ ,
\]

\[
b_2 = \ -\frac{1}{560 X^2 (1+X)^4} \times
   \bigg(24X-60 X^2+1720 X^3+5136 X^4+4768 X^5+1435 X^6
\]
\[
  -24 (1+X)^2 (1-4 X-18 X^2-12 X^3) \log[1+X] +288 X^2 (1+X)^4\, \text{PolyLog}[2,\frac{X}{1+X}]\bigg) \ .
\]
\end{widetext}
where $X=X_{\small DWP}$.

Again, the full two-loop correction takes an amazingly simple form,
\be
\label{B1DWP}
   B_1^{(DWP)}  \ = \ -\,X_{\small DWP}\,\frac{\,(4+3 \,X_{\small DWP})}{(1 + X_{\small DWP})^2} \ .
\ee
c.f. (\ref{B1AHO}), and all transcendental contributions are again cancelled out,  
and the answer is the meromorphic function of $X_{\small DWP}$ {\it only} \cite{Escobar-Ruiz:2016aqv}.

\subsection{SGP}

In this Section we consider the {\em sine-Gordon potential} (\ref{SGP})
\begin{equation*}
 V \ = \ \frac{1}{g^2}(1-\cos(g\,x))\ =\ \frac{1}{g^2}\,(1 - \cos u)\ ,\ u = g\,x\ ,
\end{equation*}
with infinite number of degenerate vacua situated periodically in $x$.

For this potential, the flucton at $T=0$ takes the form

\[
  x_{\text{flucton}}(\tau) \ =\ \frac{4\,\mbox{arccot} \bigg[ (\cosh \tau-\sinh \tau)\,\cot(\frac{g\,x_0}{4}) \bigg]}{g}  \ .
\]
The classical flucton action gives
\[
S_E[x_{\text{flucton}}]  \ =\  \frac{16\,\sin^2(\frac{g\,x_0}{4})}{g^2} \ .
\]

For the SGP, we introduce the variable

\be
\label{XSGP}
   {X_{\small SGP}}(x)\ = \ \frac{g\,x}{4}\ =\ \frac{u}{4}\ ,\ u=g x\ .\ ,
\ee
c.f. (\ref{XAHO}), (\ref{XDWP}).

Standard construction yields the following Green function

\begin{widetext}
\[
 G(\tau_1,\,\tau_2)\ =\  \frac{1}{8\,\big(\cosh(\tau_1) +\cos(2\,{X})\,\sinh(\tau_1) \big)}\times
  \frac{1}{(\cosh(\tau_2) +\cos(2\,{X_3})\,\sinh(\tau_2))}\times
\]
\[
   \bigg[ 2\,(\tau_1+\tau_2 - |\tau_2-\tau_1|)\,\sin^2(2\,{X})\ +\ 8\,\cos(2\,{X})\,\sinh^2(\frac{1}{2}(\tau_1+\tau_2 - |\tau_2-\tau_1|))
\]
\be
   +\ (3+\cos(4\,{X}))\,\sinh(\tau_1+\tau_2 - |\tau_2-\tau_1|)\ \bigg]\ ,
\ee
\end{widetext}
for $\tau_1\ ,\ \tau_2 > 0$, here $X=X_{\small SGP}$.

In this case, the ``vertex"
\begin{widetext}
\[
 \frac{\pa V_{\text{flucton}}(\tau)}{\pa X_{\small SGP}}\ =\ -\frac{16 e^{2 \tau } \sec (X_{\small SGP}){}^2 \tan(X_{\small SGP}) (e^{2 \tau }-\tan (X_{\small SGP}){}^2)}{(e^{2 \tau }+\tan (X_{\small SGP}){}^2){}^3}
 \ ,
\]
\end{widetext}
and direct evaluation of (\ref{relation}) leads to the amazingly simply expression
\be
\label{detSGP}
 \frac{\pa \log \text{Det}\,({O_{\text{flucton}} })}{\pa X_{\small SGP}}\ =\
 4 \, \tan[ {X_{\small SGP}}] \ ,
\ee
c.f. (\ref{detAHO}), (\ref{detDWP}).

As for the three two-loop Feynman diagrams shown in Fig.\ref{2loopplot} we are able to calculate analytically the diagram $a$ only,
\begin{widetext}
\[
a \ = \ -\text{Re}\bigg[\frac{1}{640} \bigg(5\ -\ 2 \sec^2(X_{\small SGP})\ +\ \sec^4(X_{\small SGP})
\]
\be
\label{aSGP}
  -\ 8\, \text{PolyLog}[2,-\tan^2(X_{\small SGP})]\ +\
  8\, [\csc^2(X_{\small SGP}) - \sec^2(X_{\small SGP})] \log[\cos(X_{\small SGP})]\bigg) \bigg] \ ,
\ee
\end{widetext}
here $\text{Re}[x]$ denotes the real part of $x$. Irrational contributions occur again (see the 2nd line of (\ref{aSGP})) like at AHO (\ref{aAHO}) and DWP (\ref{aDWP}) cases. In no way we were able to calculate $b_{1,2}$ contributions analytically. However, these two-dimensional Feynman integrals we can calculate numerically. We make two assumptions: (i) all irrational contributions cancel in the sum $B_1=a+b_1+b_2$,
see the 2nd line of (\ref{aSGP}), and (ii) the sum $B_1=a+b_1+b_2$ is given by a polynomial in $\sec^2(X)$ of degree two,
\[
    A - D \sec^2(X) + C \sec^4(X) \ ,
\]
c.f. (\ref{aSGP}), the 1st line with coefficients $A,D,C$. Based on these assumptions we fit the numerical data of $B_1$ and find that these coefficients $A,D,C$ can be calculated explicitly. It leads to a very simple expression for the two-loop correction
\be
\label{B1SGP}
B_1^{(SGP)}\ = \ -\frac{g^2}{16}\,\tan^2(X_{\small SGP}) \ .
\ee
Eventually, this expression is verified numerically. Later on this result will be derived in quantum mechanics using the generalized Bloch equation.

\subsection{The results summarized}

\label{Discussion}

The combined results from Sections C, D, E show that the expansion (\ref{Pfluc}) of the phase of the ground state function
for AHO, DWP and SGP is given explicitly by

\begin{widetext}
\begin{equation}
\begin{aligned}
\label{eqn_results}
&  2\,\phi^{(AHO)}(g,x) \ = \  {\frac{2}{3}\frac{{(1+g^2\,x^2)}^{\frac{3}{2}}-1}{g^2}} +
    {\log\bigg[\frac{(1+g^2\,x^2+\sqrt{1+g^2\,x^2})}{2}\bigg]  }  \\
&  + g^2\frac{(1-\sqrt{1+g^2\,x^2}) \left(5+16 \sqrt{1+g^2\,x^2}+25 \left(1+g^2\,x^2\right)+17
    \left(1+g^2\,x^2\right)^{\frac{3}{2}}\right)}{12(1+g^2\,x^2)^{\frac{3}{2}} \left(1+\sqrt{1+g^2\,x^2}\right)}\ +\ \ldots \\
& 2\,\phi^{(DWP)} (g,x)\ =\
     \frac{1}{g^2}\bigg((g\,x)^2 + \frac{2\,{(g\,x)}^3}{3}\bigg) \ +\  2\,\log(1 + g\,x)
     \ -\ {g^2\,\frac{(g\,x)\,(4+3\,g\,x)}{{(1+g\,x)}^2}}\ +\ \ldots  \\
&  2\,\phi^{(SGP)} (g,x) \ = \ {\frac{16}{g^2}\sin^2(\frac{g\,x}{4})}\ +\ 2\,\log[\cos(\frac{g\,x}{4})]
    \ -\ \frac{g^2}{16}\,\tan^2(\frac{g\,x}{4})\ +\ \ldots
\end{aligned}
\end{equation}
\end{widetext}
which is the Laurant expansion in powers of $g^2$ if the variable $u=g\,x$ is introduced.

\section{Re-derivation of the loop expansion from the Schr\"odinger  equation}

Our ultimate aim remains a generalization of the semiclassical theory to QFT's, thus, all our quantum-mechanical examples should be written as anharmonic perturbation of the harmonic oscillator,
\[
                 V\ =\ \frac{\tilde V(g x)}{g^2}\ ,
\]
see (\ref{potential}), where $\tilde V$ has a minimum at $x=0$ and it always starts from quadratic terms.
Classical vacuum energy is always taken to be zero, $V(0)=0$. Note that $\hbar$ was put to one, as it is traditionally done in QFT's. The  semiclassical expansion is done in powers
of  small coupling $g$ instead of the powers of $\hbar$.
It should be in the agreement with the so called non-perturbative normalization of the
non-Abelian gauge theory, in where the coupling appears $only$ in front of the action.

However, in quantum-mechanical setting traditional units are different, and in the next subsection
we show how one can reformulate these results as an expansion in the powers of the Planck constant
$\hbar$.

\subsection{Quantum-mechanical meaning of the loop expansion: generalized Bloch equation}

In order to clarify the meaning of the semiclassical loop expansion in quantum mechanics,
it is convenient to change notations as follows.
Taking the AHO potential (\ref{AHO}) as the example, we remind the units used.
The mass of the particle $m=1$, the frequency of the near-minimum oscillations $\om=1$ and the Planck constant $\hbar=1$ are all put to unity. Now we want to restore $\hbar$ in the exponent of the quantum
(statistical) weight $exp(-S_E/\hbar)$, and make a shift to the ``classical coordinate" $u=g x$. Now the Euclidean action looks as follows
\be
\label{SE-AHO}
   S_E\ =\ \frac{1}{\hbar g^2} \int d \tau
                       \left( \frac{\dot u^2}{2} + \frac{u^2 (1+u^2)}{2} \right) \ .
\ee
In other words, we have selected different unit of length, eliminating the parameter in the nonlinear, quartic term. The coupling constant now appears only together with the Planck constant. So, one can put
it to one, $g=1$, using only $\hbar$ as a parameter of the loop semiclassical expansion. The classical equation of motion does $not$ depend on it, and so is the flucton solution itself. Furthermore,
the Green function -- inverting the  operator of quantum fluctuations around the flucton -- depends on the classical coordinate $u=g\,x$ but does $not$ depend on $\hbar$. Similar consideration can be made for a general potential (\ref{potential}) and we arrive at the Euclidean action
\be
\label{SE-AHO2}
   S_E\ =\ \frac{1}{\hbar g^2} \int d \tau
                       \left( \frac{\dot u^2}{2} + {\tilde V}(u) \right) \ .
\ee

The parameters of the problem can thus be defined as (i) the quantum parameter $\hbar g^2$ (or just $\hbar$ for the $g=1$ choice) and (ii) the classical coordinate  location $u_0=g \, x_0$ under consideration.
The loop expansion of the semiclassical theory we discuss is therefore redefined
as just expansion in powers of $\hbar g^2$, starting from the classical term $O(1/\hbar g^2)$, the determinant    $O((\hbar g^2)^{0})$, the two-loop diagrams  $O((\hbar g^2)^{1})$, and so on.
For $g=1$ the loop expansion appears as the Laurant expansion in $\hbar$.
Naturally, its validity is expected when $S_{flucton}/(\hbar g^2) \gg 1$, thus, at small $\hbar \ll 1$ and/or sufficiently large $u_0$. Below we quantify the accuracy of this expansion in details.

\subsection{Iterative solution of the Schr\"odinger  equation}

\label{sec_iterative}

In this section we re-derive first three terms of the loop expansion (\ref{Pfluc}), and derive
one more term, based on quantum mechanics, employing the usual Schr\"odinger equation for the wave
function. Let us stress, that this is the only part of our program which cannot be generalized to QFT straightforwardly, at least, so far.
It allows us to cross-check of the results obtained in loop expansion.

The first step is standard, we proceed from the Schr\"odinger equation on the wave function to that
of its logarithmic derivative, which eliminates the overall normalization constant from consideration.
The second step is that we extract one power of coordinate from the function
\be
x\,  z(g\,x)\ =\ - \frac{\psi'(x)}{\psi(x)} \ .
\label{eqn_log_der}
\ee
It reflects the fact that since it is assumed the original potential (\ref{potential}) has minimum at $x=0$ the logarithmic derivative of wavefunction (the derivative of the phase) has to vanish at $x=0$.
Substituting it to the Schr\"odinger  equation

\[
\bigg(\,-\frac{1}{2}\frac{d^2}{dx^2} + V(x)\,\bigg)\,\psi(x) \ = \ E\,\psi(x)\ ,
\]
where the Planck constant is placed equal to one, $\hbar=1$, one gets the following equation on $z(g\,x)$
\be
\label{Riccati}
  \hskip -0.3cm g\,x \, z'(g\,x) +z(g\,x) -x^2 z(g\,x)^2\ =\ 2\,E - \frac{2}{g^2} \ \tilde V(g\,x)\,.
\ee
Now we redefine the coordinate
$u=g\, x$  and obtain the form of the equation we will be solving,
\be
  \hskip -0.3cm  g^2 u z'(u)\ +\ g^2 z(u)\ -\ u^2 z(u)^2\ =\ 2\, g^2\, E - 2\,\tilde V(u)\, .
\label{Riccati-Bloch}
\ee
We call this equation {\it the generalized Bloch equation}. Here $z(u)$ has a meaning of {\it reduced} logarithmic derivative. Now we proceed to solving the equation (\ref{Riccati-Bloch}).

\subsection{Weak coupling expansions}

\subsubsection{Riccati-type equation}

Let us now re-introduce logarithmic derivation
\[
    y(x)\ =\ - \frac{\psi'(x)}{\psi(x)}
\]
and write a standard Riccati equation for the potential (\ref{potential}),
\be
\label{Ricc-non-lin}
    y' - y^2\ =\ 2\,E\, -\, x^2 - 2 a_3\, g x^3 - 2 a_4\, g^2 x^4 - \ldots\ ,
\ee
cf. (\ref{Riccati}), instead of the Schr\"odinger equation, as in {\it non-linearization procedure} \cite{Turbiner:1979}, \cite{Turbiner:1984}.
Now we develop a perturbation theory in powers of $g$,
\[
    E \ =\ \sum_0^{\infty} e_m g^m\ ,\ y(x)\ =\ \sum_0^{\infty} y_m (x) g^m\ .
\]

It is evident that all $e_m$ with odd $m$ should vanish, see (\ref{energy}),
$e_{2k+1}=0, \ k=0,1,2, \ldots $. Unperturbed solution of (\ref{Ricc-non-lin}) at $g=0$ is equal to
$E_0\ =\ e_0\ =\ \frac{1}{2}$ and $y_0\ =\ x$.


The equation for the correction of the order $g$ reads,
\[
     y'_1 - 2 x y_1\ =\ - 2 a_3\, x^3\ ,
\]
with solution
\[
      y_1\ =\ a_3 x^2 + a_3\ ,\ e_1=0 \ ,
\]


The equation for the correction of the order $g^2$,
\[
     y'_2 - 2 x y_2\ =\ e_2 - 2 a_4\, x^4 + y_1^2\ =\ e_2 + a_3^2 - (2 a_4 - a_3^2)\, x^4 + 2 a_3^2 x^2 \ ,
\]
with solution

\[-
      y_2\ =\ ( a_4 - a_3^2/2) x^3 + a_3 x^2 + a x + a\ ,\ e_1=0 \ ,
\]


In general, the equation for $m-$th correction has the form


\[
   y'_m - 2 y_0 y_m\ =\ e_m - Q_m - a_m x^m\ ,
\]
\newpage

\noindent
where $Q_m = - \sum_{p=1}^{m-1} y_p y_{m-p}$ for $m>1$ play a role of effective perturbation potential: it is made from previous iterations. It can be easily demonstrated that {\it $m-$the correction $y_m(x)$ is a finite order polynomial in $x$} and, in principle, it can be found by algebraic means,
\be
\label{ym}
   y_m(x)\ =\ A^{(m)}_{m-1}\,x^{m-1}+\ldots + A^{(m)}_{k}\,x^{k}+\ldots + A^{(m)}_{0} .
\ee
A straightforward analysis leads to the remarkable property
\[
  A^{(m)}_{k} \sim \frac{m!}{k!} \ ,
\]
see \cite{Turbiner:1984}. In general, $A^{(m)}_{k}$ look as generalized Catalan numbers. Hence, the coefficient $A^{(m)}_{k}$ at fixed $k \sim m$ defines the convergent series in $m$, while at small fixed $k$ the series is usually divergent. In particular,
\[
   A^{(m)}_{1}\ =\ e_m\ .
\]

Let us change variable in (\ref{ym}),
\[
       A^{(m)}_{k} \rar A^{(m)}_{m-k}\ ,\ k=1,2,\ldots (m-1)\ .
\]
It is natural to introduce the generating function
\[
      {\tilde y}_{k}(x; \{ a \}) \ =\ \sum_{m=k}^{\infty} g^m\, A^{(m)}_{m-k}(\{ a \})\, x^{m-k}\ .
\]

If the potential (\ref{potential}) is a polynomial, several leading generating functions can be found explicitly, at $k=1,2,3,\ldots$. E.g. for AHO ($a_3=0, a_4=1$ and $a_k=0, k=5,6,\ldots$),
\[
     {\tilde y}_0\ =\  \sum_{m=0}^{\infty} g^m\, A^{(m)}_{m}\, x^{m}\ =\ x\,(1 + g^2\,x^2)^{1/2} \ ,
\]
see \cite{Turbiner:1984}, c.f. (\ref{eqn_results}).
It can be immediately recognized as the classical momentum at zero energy! Hence, the sum of leading terms of the corrections $y_m(x), m=0,1,2\cdots$ at $x \rar \infty$ is the classical momentum at zero energy:
it reminds the leading log approximation in QFT.
In the same way one can calculate the sum of next-to-leading terms of the corrections $y_m(x), m=1,2\cdots$ at $x \rar \infty$,
\[
     {\tilde y}_1\ =\  \sum_{m=1}^{\infty} g^m\, A^{(m)}_{m-1}\, x^{m-1}\ =\
\]
\[
     g^2 x\, \frac{1 + \frac{1}{2\sqrt{1+g^2\,x^2}}}{(1+g^2\,x^2+\sqrt{1+g^2\,x^2})} \ ,
\]
see \cite{Turbiner:1984}, c.f. (\ref{eqn_results}), which is the logarithmic derivative of the determinant!
Hence, the sum of sub-leading (next-after-leading) terms of the corrections $y_m(x), m=0,1,2\cdots$ at $x \rar \infty$
is the logarithmic derivative of the determinant: it reminds the next-to-leading log approximation in QFT.
We can move even further and calculate next-after-next-to-leading terms in the corrections
$y_m(x), m=1,2\cdots$ at $x \rar \infty$ and then sum them up and discover that
\[
     {\tilde y}_2\ =\  \sum_{m=2}^{\infty} g^m\, A^{(m)}_{m-2}\, x^{m-2}\ =\ \frac{d B_1}{dx}
\]
see \cite{Turbiner:1984}, c.f. (\ref{eqn_results}). Thus, the result occurs as derivative of two-loop contribution
(\ref{loopcorr}), see (\ref{B1AHO}), where $x=X_{AHO}$. It can be checked, see below,
that ${\tilde y}_3$ is the first derivative of three-loop contribution (\ref{loopcorr}).

The property that the first three generating functions ${\tilde y}_{0,1,2}$ are first derivatives of
the first three terms in loop expansion holds for DWP.
In general, the expansion in such generating functions,
\[
   y\ =\ {\tilde y}_0 + {\tilde y}_1+ {\tilde y}_2\ +\ \ldots
\]
is a {\it new} semiclassical expansion, resembling e.g. the  logarithmic approximations of QFT.

\subsubsection{Generalized Bloch equation case}

Calculations of the loop expansion for the ground state wave function phase performed earlier for AHO, DWP and SGP show that this expansion looks like a perturbation series in $g^2$ (starting from $1/g^2$ term) if classical coordinate $u=g\,x$ is introduced. It is natural to construct this perturbation theory for phase in generalized Bloch equation (\ref{Riccati-Bloch}) and compare with the loop calculation.

Solving (\ref{Riccati-Bloch}) iteratively we generate the flucton loop expansion.
Let us define the series
\be
z(u)\ =\ \sum_{n=0} g^{2n} z_n(u)\ ,
\ee
which corresponds to perturbative solution of this equation in powers of $g^2$ with
\[
        E\ =\ \sum_{n=0} g^{2n} E_n\ ,
\]
given by standard perturbation theory in $g^2$.

In the zeroth order, in which all terms proportional to the coupling are ignored, the
equation is very simple
\be
   -\, u^2 \,z_0(u)^2\ =\ - 2\tilde V(u)\ ,
\ee
leading to
\be
\label{z0}
     z_0(u)\ =\ \frac{\sqrt{2\,\tilde V(u)}}{u}\ .
\ee
This result $(u\, z_0)$ is, in fact, the classical momentum at zero energy, and therefore,
as one returns to the wave function, the zeroth order term gives the well known semiclassical action.
So, at this stage, the result is well known $\psi \sim exp(-\int^x p(x') dx')$ but at zero energy.
It can be immediately checked that classical flucton action for AHO, DWP and SGP, see (\ref{eqn_results}),
is nothing but the semiclassical action at zero energy, $\int u\, z_0(u) du$ with $z_0(u)$ given by (\ref{z0}).

Moving to the next term of the expansion, one finds the following equation $O(g^2)$ for it
\be
\label{Ez1}
u z_0'(u)\ +\ z_0(u)\ -\ 2 \, u^2 z_0(u) z_1(u)\ =\ 2\,E_0\ .
\ee
Note here, that the equation involves the known function $z_0$ of the previous order,
and $z_1$ just appears linearly. The similar feature takes place in all orders!

Important point of the correct procedure is that the energy needs to be used in the form of perturbative
expansion in powers of $g^2$ as well but for the original potential (\ref{potential}),
\be
\label{E-PT}
     E=\sum_{n=0} g^{2n} E_n\ .
\ee
The zeroth order potential is of the harmonic oscillator, so $2\,E_0=1$. Hence, the first correction
\be
\label{z1}
   z_1(u)\ =\ \frac{u z_0'(u)\ +\ z_0(u)\ -\ 1}{2 \, u^2 z_0(u)}\ .
\ee
It can be immediately checked that that the logarithm of determinant $\log \text{Det}\,(O_{\text{flucton}})$ (\ref{relation}) for all three potentials AHO, DWP and SGP, see (\ref{eqn_results}) is nothing but $\int u\, z_1(u) du$ with $z_1(u)$ given by (\ref{z1}). Thus, in a very simple way we calculated determinant, $\log \text{Det}\,(O_{\text{flucton}})$ explicitly (!) in closed analytic form for the general potential $V(x)$ (\ref{potential}).
Or, in other words, we calculated explicitly the one-loop diagram of Fig. \ref{fig_oneloop}.
This result,  written in terms of classical flucton action and its derivatives,
is one of the central results of this paper.

Moving to the next term of the expansion, one finds the following equation $O(g^4)$ for it
\be
\label{Ez2}
u z_1'(u) + z_1(u) - u^2 z^2_1(u) - 2u^2 z_0(u) z_2(u) = 2 E_1\ .
\ee
Note here, that the equation involves the known functions of the previous orders $z_{0,1}$
non-trivially, but the new function  $z_2$ appears only linearly. This feature is generic,
repeated in each order: so there is no difficulty to find new corrections.

The perturbative coefficient $E_1$ is the perturbative correction $\sim g^2$ to the ground state energy in the quartic part of the original potential (\ref{potential}),
\[
  V(x)\ =\ \frac{1}{2}\, x^2 + a_3\, g x^3 + a_4\, g^2 x^4\ ,
\]
which can be easily found explicitly e.g. in non-linearization procedure \cite{Turbiner:1984},
\[
    2 E_1\ =\ \frac{3}{2}\,a_4\,-\,\frac{11}{4}\, a_3^2\ .
\]
Solving (\ref{Ez2}), we find the second correction
\be
\label{z2}
   z_2(u)\ =\ \frac{u z_1'(u)\ +\ z_1(u)\ -\ u^2 z^2_1(u)\ -\ 2\,E_1}{2 \, u^2 z_0(u)}\ ,
\ee
which defines the two-loop contribution $B_1$. It can be immediately checked that for AHO, DWP and SGP, see (\ref{eqn_results}) is nothing but $\int u\, z_2(u) du$ with $z_2(u)$ given by (\ref{z2}). Thus, in a very simple way we calculated two-loop contribution explicitly (!) in closed analytic form
for the general potential $V(x)$ (\ref{potential}).
Or, in other words, we calculated sum of three two-loop diagrams on Fig. \ref{2loopplot}, weighted with symmetry factors, explicitly. For AHO, DWP and SGP potentials this sum $\int u\, z_2(u) du$ with $z_2(u)$ does {\it not} contain transcendental contributions. It must be emphasized again that it was difficult to guess that such a result can exist in such a generality.

Moving to the next term of the expansion, one finds the following equation $O(g^6)$ for it
\[
   u z_2'(u) + z_2(u) - 2 u^2 z_1(u) z_2(u) - 2u^2 z_0(u) z_3(u)
\]
\be
\label{Ez3}
   =\ 2\, E_2\ .
\ee
Note here, that the equation involves the known functions of the previous orders $z_{0,1,2}$,
and again $z_3$ appears linearly. The perturbative coefficient $E_2$ is the perturbative
correction $\sim g^4$ to the ground state energy in the sextic part,
\[
  V(x)\ =\ \frac{1}{2}\, x^2 + a_3\, g x^3 + a_4\, g^2 x^4 + a_5\, g x^5 + a_6\, g^2 x^6\ ,
\]
of the general  potential (\ref{potential}), which can be straightforwardly found e.g. in non-linearization procedure \cite{Turbiner:1984}. Eventually, the third correction
\be
\label{z3}
   z_3(u)\, =\, \frac{u z_2'(u)\, +\, z_2(u)\, -\, 2 u^2 z_1(u) z_2(u)\, -\, 2\,E_2}{2 \, u^2 z_0(u)}\ ,
\ee
defines the three-loop contribution $B_2$ analytically. In the Feynman diagram technique (the flucton formalism) it corresponds in general to sum of 15 three-loop diagrams on Fig. 2 in \cite{Escobar-Ruiz:2015rfa} weighted with symmetry factors. No single diagram we were able to calculate analytically for AHO, DWP and SGP potentials yet. Needless to say that next iterations will provide higher loop contributions in the same straightforward way.

\subsection{AHO: three-loop correction}

Starting from two-loop correction $z_2(u)$ the details of specific example become relevant,
since one needs a concrete value for $E_1$. So, from this point on, we present for
the AHO case (\ref{AHO}) one more term, three-loop correction for AHO. In order to do it we repeat consideration of the previous section in brief for the case of AHO.

For convenience, we introduce a new variable $s=u^2$. Then the generalized Bloch equation (\ref{Riccati-Bloch}) takes the form
\be
\label{RB-AHO}
 2\,g^2\,s\,z'(s) + g^2\,z(s)  - s\,z(s)^2 \ = \ 2 E\,g^2 - s(1+s) ,
\ee
where $s \in [0,\infty)$. At zero order $(g^2)^0$ we have the equation
\be
\label{z0-AHO}
z_0(s)^2 \ = \ (1+s) \ ,
\ee
cf. (\ref{z0}) then for normalizability of the wave function it is required to take the positive solution $z_0(s)= \sqrt{1+s}$. The equation to the next order $g^2$ is given by
\be
2\,s\,z_0'(s) + z_0(s)\,(\,1-2\,s\,z_1(s)\,)  \ = \ 2\,E_0 \ ,
\ee
from which it follows that
\[
z_1(s) \ = \ \frac{1 + 2\,s - 2\, E_0\, \sqrt{1+s}}{2 s (1+s)} \ ,\ 2\, E_0=1\ ,
\]
Note that the condition that the function $z_1(s)$ is not singular at the origin, $s=0$ also implies
$2\, E_0=1$. Now, vanishing of the coefficient in (\ref{RB-AHO}) of order of $g^4$ leads to the equation for the second correction $z_2(s)$ , which is equal to
\begin{widetext}
\be
z_2(s) \ = \ \frac{- E_1 + z_1 - s z_1^2 + 2 s z_1'}{2 s z_0} \
= \ \frac{4 \,(-1+\sqrt{1+s})+s(-7-8 \,s+8 \sqrt{1+s})-4 s (1+s)^2 E_1}{8\, s^2\, (1+s)^{5/2}}\ ,
\ee
\end{widetext}
where $z_1(s)$ is already known and $E_1=\frac{3}{4}$ is well-known first energy correction to the AHO ground state. At small $s \rightarrow 0$ we obtain
\begin{widetext}
\be
z_2(s) \approx
 \frac{3-4 E_1}{8 s}+\frac{1}{4} \left(-6+E_1\right)-\frac{3}{64} s \left(-63+4 E_1\right) + \ldots  \ .
\ee
\end{widetext}
The value $E_1=\frac{3}{4}$ leads to disappearance of the first (singular) term in this expansion.
Similarly, we obtain
\begin{widetext}
\be
z_3(s)\ = \ \frac{1}{32 \,\si^8 \,(1+\si)^3} \bigg[60 + 230 \si + 346 \si^2\ +
%
  270 \si^3 + 150 \si^4+108 \si^5+84 \si^6+63 \si^7 + 21 \si^8 \bigg] \ ,
\ee
\end{widetext}
where
$E_2=-\frac{21}{16}$, and $\si = \sqrt{1+s}$. Of course, in the variable $\si$ which is nothing but
$z_0(s)$ (\ref{z0-AHO}) it can be easily seen that all corrections $z_n(s)$ are meromorphic functions, no transcendental terms occur. From (\ref{eqn_log_der}) we immediately make the identifications
\begin{equation}
\begin{aligned}
 &  z_0(s) \ = \ g^2\,  \pa_s S_{\text flucton}  \ ,\\
 &  z_1(s) \ = \  \pa_s \bigg(   \frac{1}{2} \log \text{Det(flucton)}  \bigg) \ ,\\
 &  z_2(s) \ = \  g^{-2} \, \pa_s (\text{two-loop}) \ , \\
 &  z_n(s) \ = \ g^{-2(n-1)} \, \pa_s (\text{n-loop}) \ .
\end{aligned}
\end{equation}

\bigskip

\subsection{ADWP: classical action and one-, two-, three-loop corrections}

Finally, within the iteration method for the generalized Bloch equation (\ref{Riccati-Bloch}), we consider the asymmetric double-well potential (ADWP), see (\ref{ADWP}),
\[
                 V\ =\ \frac{1}{2} x^2 (1 + 2\,t\,g\, x + g^2\,x^2)\ ,\ t \in [0,1]\ ,
\]
which is, in fact, general quartic potential.

\bigskip

\begin{figure}[h]
\begin{center}
\includegraphics[width=6cm]{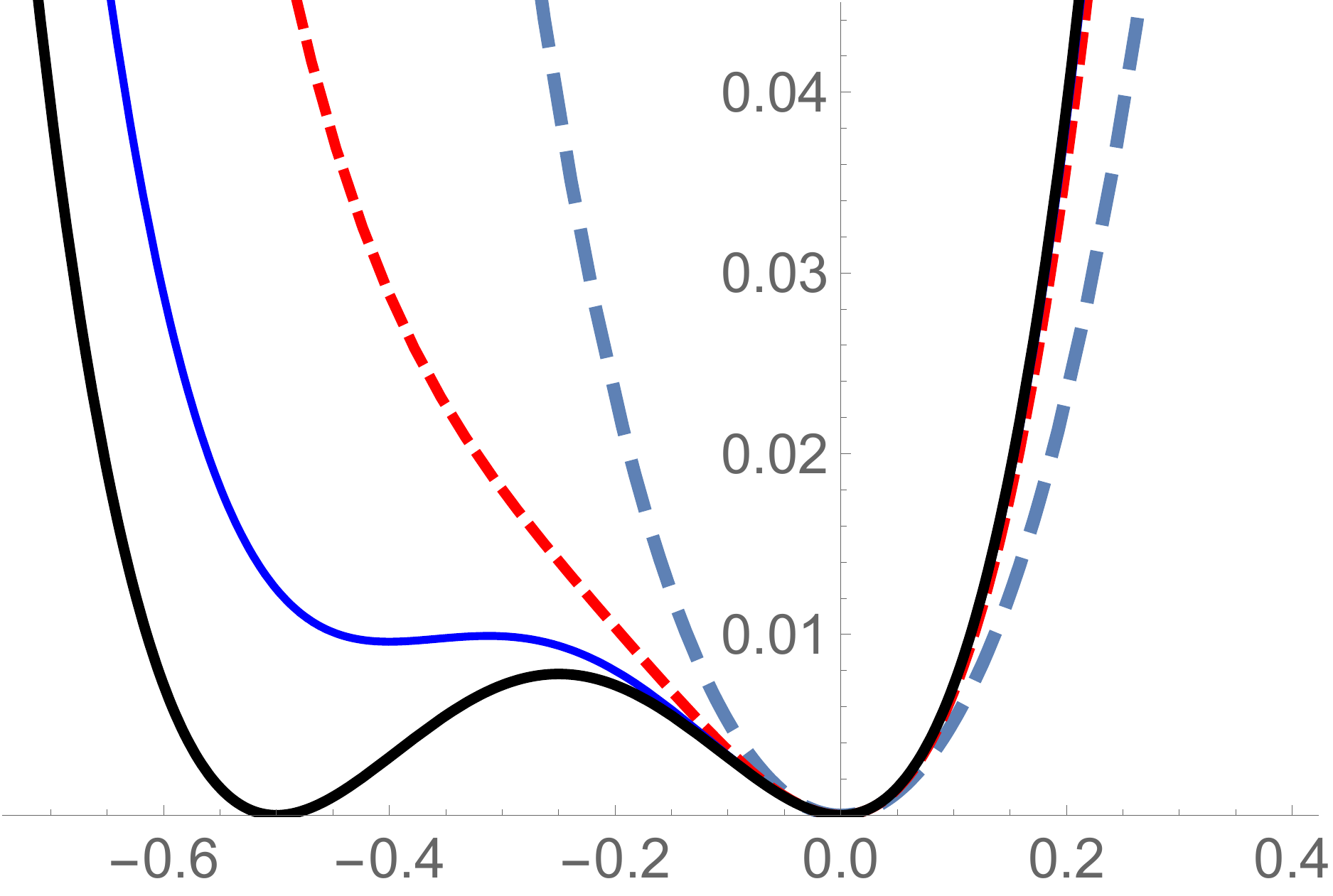}
\caption{The ADWP for \text{t}=0 (solid black), 0.8 (solid blue), 0.95 (short dashed, red), 1 (dashed grey) at the coupling $g=2$.}
\label{FigADWP}
\end{center}
\end{figure}

In this case, one potential minimum is situated at $x=0$ and $V(0)=0$, while for the second minimum (when exists) is situated to left from $x=0$ and $V(x_{min}) \geq 0$.


The generalized Bloch equation (\ref{Riccati-Bloch}) takes the form
\begin{widetext}
\be
\label{RB-ADWP}
     g^2 u z'(u) + g^2 z(u) - u^2 z(u)^2 - 2 \,E\, g^2 + u^2 (1+2 \,t \,u + u^2)\ =\  0 \ .
\ee
\end{widetext}

After a straightforward calculation, see e.g. \cite{Turbiner:1984} we find explicitly the first three coefficients of the perturbative expansion in $g^2$ of the energy (\ref{E-PT}),
\begin{widetext}
\begin{equation}
\begin{aligned}
 & E_0 \ = \ 1  \ ,  \\
 & E_1 \ = \ \frac{1}{4} \left(3 - 11 \,t^2\right)  \ , \\
 & E_2 \ = \ -\frac{3}{16} \left(7 - 114 \,t^2 + 155 \,t^4\right)  \ .
\end{aligned}
\end{equation}
\end{widetext}
These coefficients will be needed to find one-, two-, three-loop contributions in iteration method applied for (\ref{RB-ADWP}).


Zero iteration of (\ref{RB-ADWP}) gives the classical momentum at zero energy
\begin{widetext}
\[
   z_0\ =\ u \sqrt{1 + 2\,t\,u\ +\ u^2}\ =\ g\,x \sqrt{1 + 2\,t\,g\, x + g^2\,x^2}\ \equiv\ (g\,x)\,X^{(ADWP)}\ ,
\]
\end{widetext}
where for convenience we will denote hereafter $X^{(ADWP)} \equiv X_4$\,,
while the classical flucton action $\sim~\int u z_0(u)$ reads,
\begin{widetext}
\begin{equation}
\label{S-ADWP}
\begin{aligned}
 & S_{\text{flucton}}\ =\
        \frac{-2+3 t^2 + 3 t \left(1-t^2\right) \left(\text{Log}[1+t]-\text{Log} [F+X_4]\right)\ -\ 3 F\, t\, X_4\ +\ 2 \,X_4^3}{3\, g^2} \ ,
\end{aligned}
\end{equation}
\end{widetext}
where $F\ \equiv\ \sqrt{-1+t^2+X_4^2}$. As for the determinant, $\log \text{Det}\,({O_{\text{flucton}} })$ it is equal to
\begin{widetext}
\begin{equation}
\label{Det-ADWP}
\begin{aligned}
 & \frac{1}{2} \log \text{Det(flucton)} \  =\ \text{Log}\left[\frac{X_4}{2}\right]+\text{Log}\left[1+(F-t) t+X_4\right]  \ .
\end{aligned}
\end{equation}
\end{widetext}

It can be immediately checked that taking (\ref{S-ADWP}) and (\ref{Det-ADWP}) at $t=0$ we recover the results for the AHO and at $t = \pm 1$ those of the DWP. As for two-loop correction $B_1$ it takes the form

\begin{widetext}
\[
B_1^{(ADWP)}\ =\ \frac{1}{12 (F-t)^2 \left(-1+t^2\right) X_4^3}\ \bigg( 5 (1+ F t - 2 t^2 - F t^3 + t^4)
+ 6(1 - t^2) X_4 - 2(1 + F t - t^2) X_4^2
\]
\[
+\left(-17-34 F t+134 t^2+200 F t^3-285 t^4-170 F t^5+170 t^6\right) X_4^3+\left(-9+31 F t-20 t^2-33 F t^3+33 t^4\right) X_4^4
\]
\[
+\left(17-100 t^2+85 t^4\right) X_4^5\bigg)  \ .
\]
\end{widetext}
In the limits $t = \pm 1$ and $t=0$ it coincides with two-loop correction $B_1$ for DWP and AHO, respectively.

The three-loop correction $B_2$, which in the path integral formalism is given by the sum of 15 weighted (with symmetry factors) Feynman integrals (running from one-dimensional up to six-dimensional integrals), in the iterative approach to the generalized Bloch equation (\ref{Riccati-Bloch}) can be easily calculated

\begin{widetext}
\begin{equation}
\label{B2-ADWP}
B_2^{(ADWP)}=\frac{1}{64\,(F-t)^4}  \bigg[  \frac{40 \left(1+2 F t-2 t^2\right) \left(-1+t^2\right)}{ X_4^6}-\frac{40 \left(1+F t-t^2\right)}{ X_4^5})
\end{equation}
\[
+\frac{8 \left(8+8 F t-13 t^2\right)}{ X_4^4}+\frac{64}{ X_4^3}-\frac{8 \left(5-6 F t+17 t^2+22 F t^3-22 t^4\right)}{ X_4^2}+\frac{8 \left(-5+11 F t-11 t^2\right)}{ X_4}
\]
\[
+ \bigg( (207-3598 t^2+5639 t^4)(F-t)^4 + 8 \left(-3+11 t^2\right)\bigg) + 4 \left(35+426 t^2-1833 t^4+1860 t^6-3 F t \left(69-301 t^2+620 t^4\right)\right) X_4
\]

\[
 -\ 4 \left(21+406 t^2-1395 t^4+F t \left(-187+465 t^2\right)\right) X_4^3\bigg] \ .
\]
\end{widetext}
It is rather surprising, that $B_2^{(ADWP)}$ is given by so compact expression.
In particular,

\begin{widetext}
\begin{equation}
\begin{aligned}
& B_2^{(AHO)}\ = \ -\frac{40+120\,X_{AHO} + 136\,X_{AHO}^2 + 88\,X_{AHO}^3 + 80\,X_{AHO}^4 +
112\,X_{AHO}^5 - 39\,X_{AHO}^6 - 330\,X_{AHO}^7 - 207\,X_{AHO}^8}{64\, X_{AHO}^6 \,(1+X_{AHO})^2}
  \\ \\
& B_2^{(DWP)} \  = \  \ \frac{X_{DWP}\,(128 + 300\,X_{DWP} + 248\,X_{DWP}^2 + 71\,X_{DWP}^3)}{4\,(1+X_{DWP})^4}
  \  .
\end{aligned}
\end{equation}
\end{widetext}

For completeness we present the three-loop correction for the SGP

\begin{equation}
\label{B2-SGP}
  B_2^{(SGP)}\ = \ \frac{7-\left(6+\cos(X_{SGP})\right) {\sec}^4 \left(X_{SGP}\right)}{1024}   \ .
\end{equation}
Note it is of the amazingly simple form.

\subsection{Strong coupling expansion}

So far, we studied weak coupling expansion for the generalized Bloch equation (\ref{Riccati-Bloch}),
\[
z(u)\ =\ \sum_{n=0} g^{2n} z_n(u)\ ,\ E\ =\ \sum_{n=0} g^{2n} E_n\ ,
\]
which corresponds to perturbation theory in $g^2$. Now we will study the strong coupling expansion in $1/g$.
It is convenient to consider a particular potential breaking the idea of generality. We present here the results for the AHO case (\ref{AHO})
\[
                   V\ =\ \frac{1}{2}\,x^2 (1 + g^2 x^4)\ .
\]
Let us introduce as first the classical coordinate $u=g\,x$ and then as second introduce new variable $s=u^2$. Then the generalized Bloch equation (\ref{Riccati-Bloch}) takes the form (\ref{RB-AHO}),
\[
 2\,g^2\,s\,z'(s) + g^2\,z(s)  - s\,z(s)^2 \ = \ 2 E\,g^2 - s(1+s) ,
\]
where $s \in [0,\infty)$. Since we know (functionally) the strong coupling expansion for energy, see e.g. \cite{Turbiner:1988tu}, let is develop perturbation theory
\be
\label{PT-strong}
        E= g^{2/3} \sum b_n g^{-\frac{4n}{3}}\ ,\ z = g^{2/3} \sum F_n(s) g^{-\frac{4n}{3}}\ ,
\ee
where $b_n, n=0,1,2\ldots$ are strong coupling coefficients for energy, few of them are found numerically with high accuracy.
The equation for finding the zero order $O(g^{2+\frac{2}{3}})$ is of the form
\[
  2 s F_0' + F_0\ =\ b_0\ ,
\]
it does {\it not} depend on the potential explicitly and
\be
   F_0\ =\ b_0\ .
\ee
The equation for the 1st order correction $O(g^{2-\frac{2}{3}})$,
\[
  2 s F_1' + F_1 - s F_0^2\ =\ b_1\ ,
\]
which also does {\it not} depend on the potential explicitly,
\be
   F_1\ =\ \frac{b_0^2}{3} s + b_1\ .
\ee
It can be easily shown that the $n$th correction is a polynomial of degree $n$,
\[
    F_n\ =\ \alpha_0(b) s^n + \alpha_1(b) s^{n-1} + \ldots + b_n\ .
\]

\section{The accuracy of the perturbative and the semiclassical loop expansions}

In this section we address the issues of convergence and  accuracy of the
expressions derived above, comparing the terms of the expansion to each other and to
 the  wave functions
obtained numerically. In particular,
this comparison will quantify the meaning of ``large classical coordinate" $y=g\ x$ and
``small coupling" $g$.
For definiteness, we discuss those issues for the case of AHO system.

\begin{figure}[h!]
\begin{center}
\includegraphics[width=8cm]{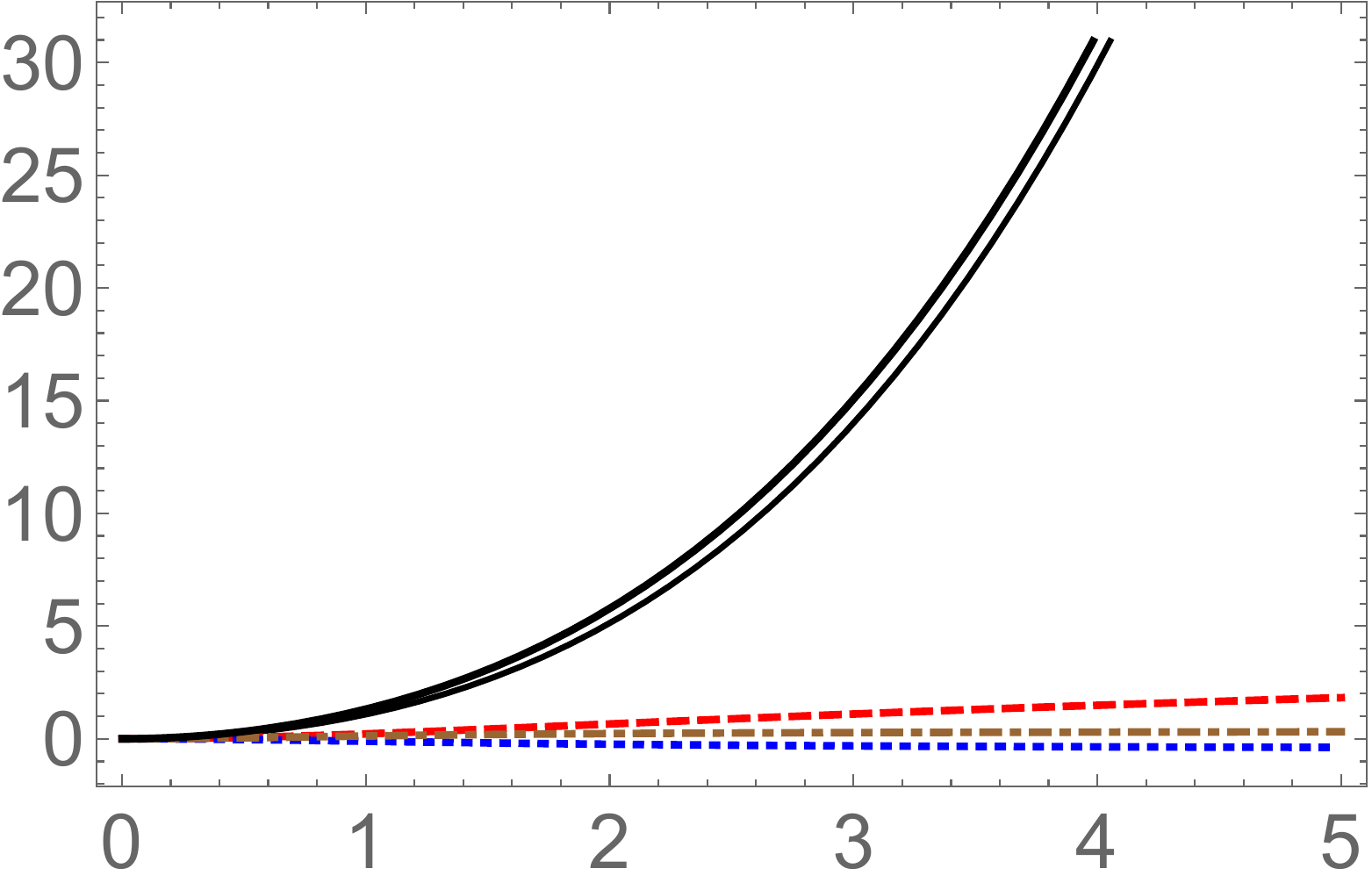}
\caption{(color online) The (double) phase
of the wave function $2\phi$
 for the anharmonic oscillator versus the coordinate $x$.
for the coupling  $g^2=1/3$. 
The (lower) thin black solid line is the leading term, corresponding to the
classical flucton action. The red dashed, blue dotted and brown
dot-dashed lines show the magnitude of the one, two and three loop corrections.
Their sum is shown by the (upper) thick black solid line.}
\label{fig_AHO_weak}
\end{center}
\end{figure}

Let us first address the issue of convergence of the perturbative series in $g$
at weak coupling. In Fig.\ref{fig_AHO_weak}
we plot four subsequent term of the expansion of the wave function phase
and their sum, for ``small"   $g^2=1/3$ coupling.
and ``large" $g^2=2$. In all cases there is clear dominance of the classical $O(g^{-2})$ term at large values of the coordinates, $x\gg 1$. This happens because only the classical term
grows with $x$. But if one excludes the leading term and compares the subsequent loop corrections themselves, the series seem to be convergent at $all$ $x$ rather well.

\begin{figure}[b]
\begin{center}
\includegraphics[width=8cm]{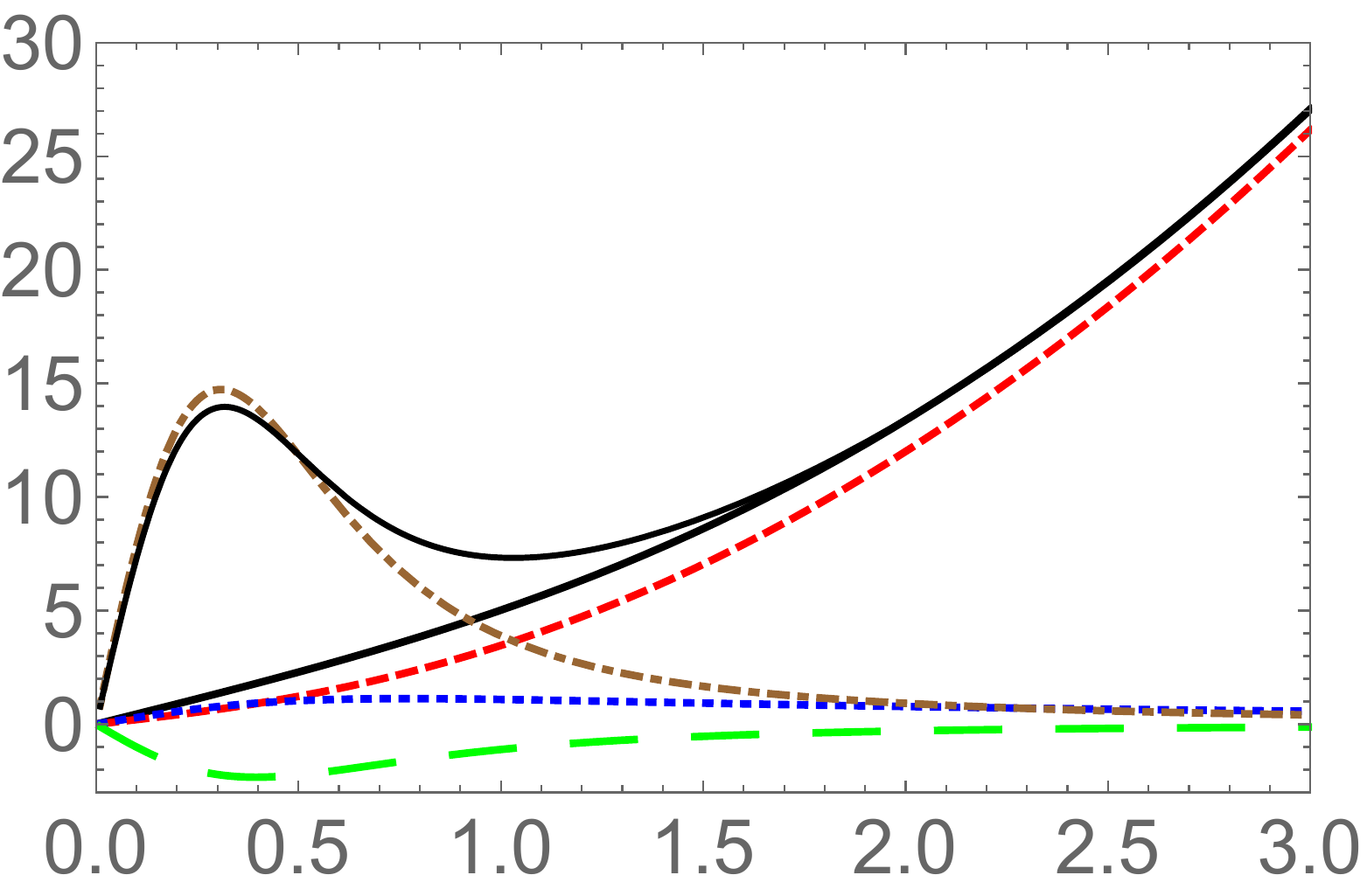}
\caption{(color online) The comparison between  the variational  wave function squared \cite{Turbiner:2010} (thick black solid line) with
the flucton loop expansion, zero to three loops, at
the coupling $g^2=2$. The red dashed, blue dotted, green long-dashed, brown dot-dashed and thin solid black lines are for classical action, 1-, 2-, 3-loop contributions and their sum, respectively.}
\label{fig_AHO_strong}
\end{center}
\end{figure}

We now proceed to the case of strong coupling, $g>1$, and ask whether the semiclassical theory is still applicable in its domain, at large values of the coordinate.
In Fig.\ref{fig_AHO_strong} we compare four terms of the loop expansion
with the extremely accurate variational wave function derived
previously by one of us \cite{Turbiner:2010}, for rather strong coupling $g^2=2$.

Two observations come from this plot. The first is that
at strong coupling the convergence at small $x<1$ is gone. Yet the second is
that in the semiclassical domain $x>1$ one can see that
higher loop corrections do in fact improve the classical result.
In fact, for  $x>1.5$ the difference between the flucton series (up to 4th term)
and the variational curve is smaller than the width of the line!

%

%

The last issue we discuss in this section is that of the overall normalization constant.
The flucton method, by construction, is designed to give the $relative$ probability to find a particle at different locations. Rather arbitrarily, we have selected in all the discussion above
the ``normalization point" to be located at the potential minimum. Indeed, our flucton
and its action are both zero, for a particle located there. So the same value  -- taken to be one -- is used for all wave functions at the maximum $x=0$.

\begin{figure}[b!]
\begin{center}
\includegraphics[width=8cm]{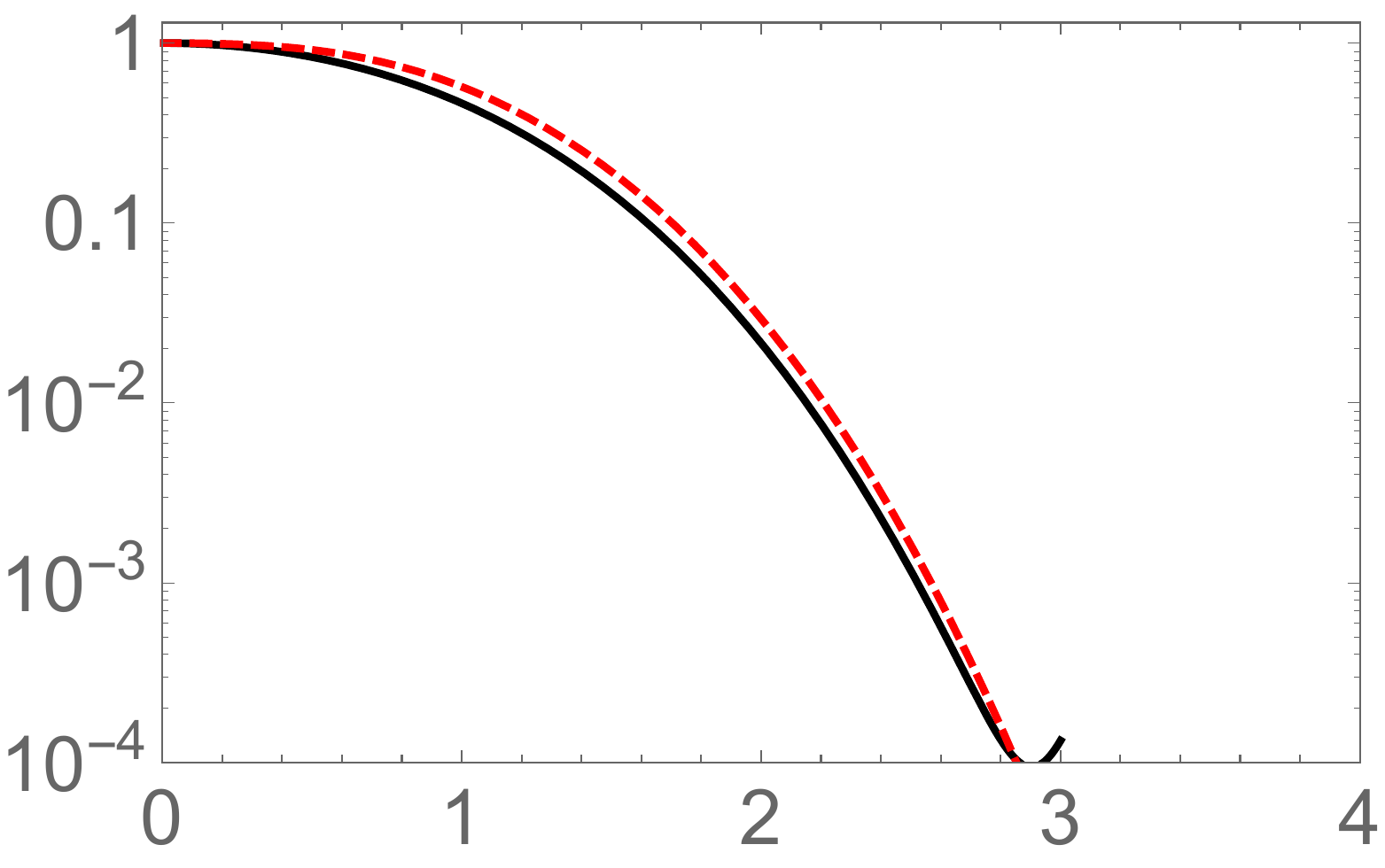}
\includegraphics[width=8cm]{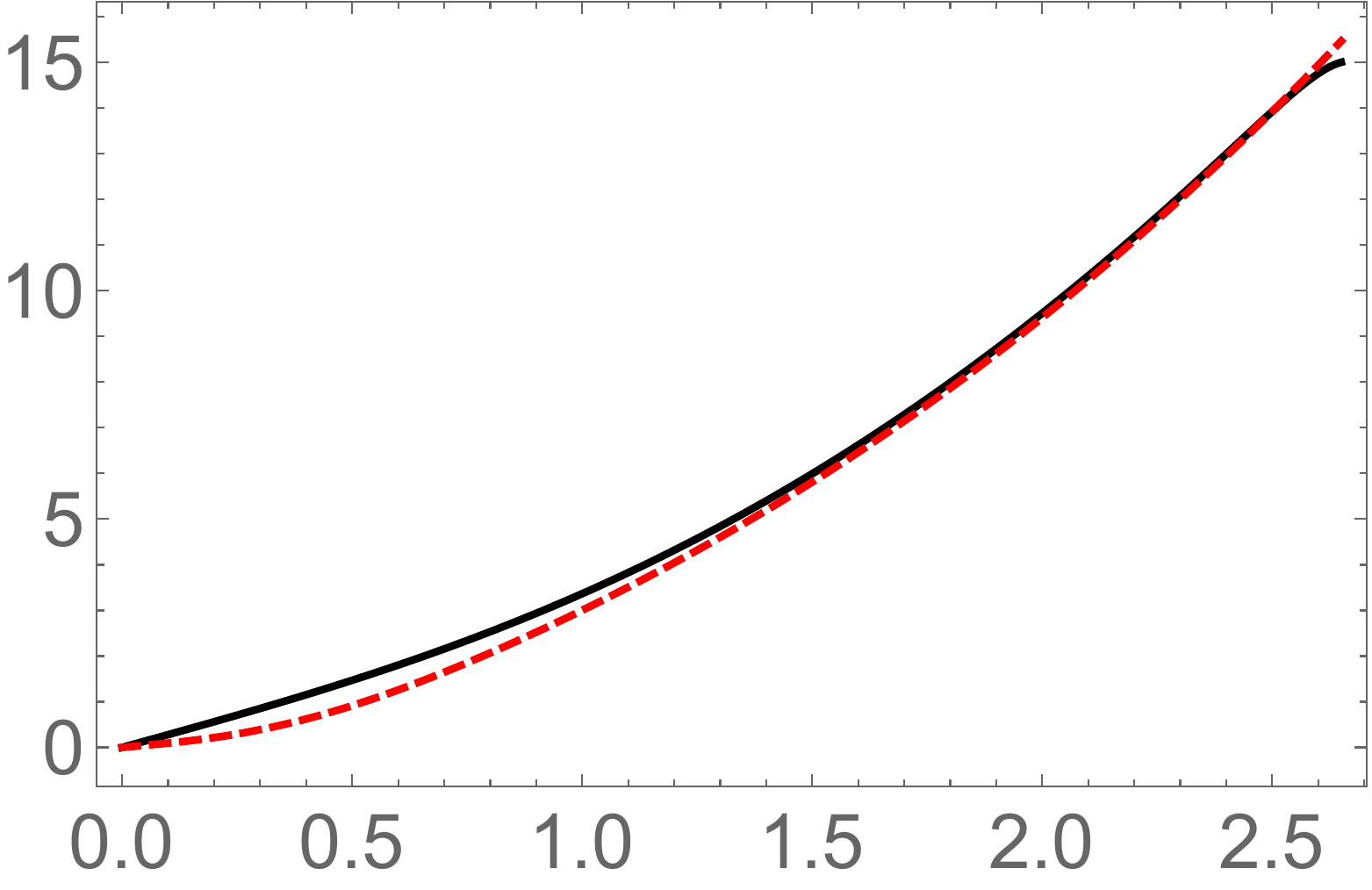}
\caption{(color online) The comparisons for the anharmonic oscillator with
the coupling $g^2=1$ of the
numerically calculated wave function (black solid lines) with the
flucton expansion (0-1-2-loops summed) shown by the red dashed lines.
The upper and lower plots show the density in the ground state and its logarithmic derivative
versus the coordinate $x$, respectively.}
\label{fig_AHO_exact}
\end{center}
\end{figure}

The upper Fig.\ref{fig_AHO_exact} shows a comparison of the semiclassical density matrix
(the sum of 3  terms)  with the exact (numerically calculated at energy $E_0 = 0.69617575$) wave function squared, for $g=1$. While such normalization is natural for the semiclasical approach used, it is
in fact inadequate, in the following sense. As it is clear from the upper plot of  Fig.\ref{fig_AHO_exact},  this normalization does not provide good description at large $x$, which is the semiclassical domain.

Such outcome is of course not unexpected. Our derivation from the Schr\"odinger  equation
in section  \ref{sec_iterative} is based on the {\em logarithmic derivative} of the wave function (\ref{eqn_log_der}), which does not depend on the normalization constant.
Therefore, a more meaningful comparison between the semiclassical expansion and the exact wave function
can be provided by the plot of the corresponding  logarithmic derivatives. Such comparison is shown in the lower plot of Fig.\ref{fig_AHO_exact}: now the agreement between the two curves is observed for $x>1$, in
the semiclassical domain. Outside it, at $x<1$, the agreement is not expected, but it is not too bad either. (Note that this figure corresponds to  the  coupling which is not small, $g^2=1$).


\section{The semiclassical expansion for potentials with multiple minima}

In general, one may think also of the potentials with $N$ minima, and ask how the flucton-based approximation for the path integral we develop should be applied in this case.

The case we start with has all  minima to be $degenerate$, corresponding
to the same energy (which then can always be put to zero). Since in this case
all of the minima can be used for ``long-time relaxation" of the flucton paths, one can think of $N\times N$
matrix of fluctons $x_{ij}^{f}(\tau)$, starting at $\tau\rightarrow -\infty$ in the $i$-th vacuum and ending at  $\tau\rightarrow \infty$ in the $j$-th one. Of course,
for a given  ``observation point" $x_0$ one only needs to consider those paths which pass through it.  The DWP  is the example of such degenerate situation, to be discussed in the subsection \ref{sec_DWP_between}.

However the problems with the  $non-degenerate$ minima, such as  in the ADWP case,
obviously cannot be treated in this way. There are no non-diagonal paths
$i\neq j$ between different maxima which can ``relax" at both ends, as those have different energy. Long-time ``relaxation" is obviously only possible at the global minima.
This situation, to be discussed in subsection \ref{sec_ADWP_complex}, require
 complexified classical paths.

\subsection{ The density between the minima,
for the symmetric double well } \label{sec_DWP_between}

So far we only discussed the ``outer region" $| x_0 |> x_{min}=1/g$ outside its two minima.
Now let us discuss the intermediate region, around the middle point $x\sim 0$.
For sufficiently small $g$ -- and thus well separated minima -- it should also be
amenable to a semiclassical treatment.

%

Following the discussion above, the double well problem should have 4 flucton paths. The $x_{11}^{f}(\tau)$ and $x_{22}^{f}(\tau)$ are fluctons we discussed, associated with say the left and the right potential minima. Their contributions generate two familiar maxima in the ground state wave function.

In the outer region  $| x_0 |> x_{min}$ the  $x_{12}^{f}(\tau)$ and $x_{21}^{f}(\tau)$ fluctons are
a combination of the $x_{11}^{f}(\tau)$ and
$x_{22}^{f}(\tau)$  fluctons, $plus$ the $instanton$ or $antiinstanton$ paths. So the actions are just
\be
S_{12}^{f}=S_{11}^f(x_0) + S_{instanton} \ .
\ee
This means that the density matrix due to the $(11)$ flucton, is just corrected by an exponentially small  and $x_0$-independent term
\be
       \psi_0^2(x_0) \sim  exp(-S_{11}^f(x_0)) \left( 1  + O(e^{-S_{instanton}}\right) )\ .
\ee

If  $| x_0 |< x_{min} $, in the inner region,
the $x_{12}^{f}(\tau)$ and $x_{21}^{f}(\tau)$ fluctons are nothing else as the $instanton$ and $antiinstanton$ paths.
Their timing can be selected so that at $\tau=0$ their value, as for all other fluctons, should be $x_0$.
Furthermore, their classical actions
\[
 S_{12}^{f}=S_{21}^{f}=\int_{-x_{min}}^{x_0} p(x') dx' +\int_{x_0}^ {x_{min}} p(x') dx'
\]
\be
     =\ \int_{-x_{min}}^{x_{min}} p(x') dx' \ ,
\ee
do not depend on the $x_0$. Therefore, their contribution to the density matrix in the inner region
is -- somewhat surprisingly -- independent on the observation point $x_0$.

It may appear strange that the flucton theory has such unusual contributions in the inner region.
since in the familiar WKB-like semiclassical theory one does not have those.
Note however, that the WKB is applied to  the wave function, while the   flucton theory
is applied to the density matrix, or its $square$  $\psi_0^2 $: the instanton terms thus come
from the product of the two semiclassical contributions in the WKB-like approaches.

To demonstrate its validity, we will use Turbiner trial function $cosh(A)$, see \cite{Turbiner:2010},
\be
cosh^2(A)=(1/2) \left(cosh(2A) +2\right)
\ee

\subsection{The density for the asymmetric double well: complex fluctons}
\label{sec_ADWP_complex}

As we already noted at the beginning of this section, when the minima of the potential are non-degenerate
there are no classical solutions going from one maximum to the other of the potential in Euclidian time
and ``relaxing" at both ends, simply because for that one needs two conflicting values of the energy.
In particular, there are no ``instanton" and ``anti-instanton" solutions available.

It was argued in  \cite{Kozcaz:2016wvy,Behtash:2015zha}  that
%
by complexification of the coordinate, $x(t)\rightarrow z(t)=x(t)+i\cdot y(t)$ and
thus by generalizing equations of motion to the so-called {\em holomorphic
Newton's equation} (still for inverted potential)
\be
\label{hNE}
    \frac{d^2 z}{dt^2}\ =\ + {\pa V \over \pa z} \ ,
\ee
one can find complex generalization of those.
Specifically, in these works it is discussed the contribution of a periodic path
with a finite action, called the ``complexified bion" (CB), which is an extension to the instanton-antiinstanton pair solutions for the symmetric potential.

For continuity of the notations, let us use the following (Euclidean time) Lagrangian
\be
 {\cal L}_{ADWP}\ =\ \frac{1}{2}\, \dot x(\tau)^2 + \frac{1}{2}\,(x(\tau)^2 - 1)^2\ +\ p\, g x(\tau)\ ,
\ee
with the asymmetry parameter $p$.  If $p$ is nonzero but small, the left and right
maxima located at $x_+$ and $x_-$  are of different height, $E_+=V(x_+)\neq E_-=V(x_-)$
\footnote{
For illustration below we will use the case $p g = 0.1$, for which these locations are
$x_+\approx -1.02412, x_-\approx 0.973994$. Note that plus and minus in our notations
do not correspond to the sign of the coordinates, but to the height of the inverted potential,
$V(x_+)=0.10122 > V(x_-)=-0.0987171$.}

Like for symmetric case discussed before, for a generic point $x_0$ in between the two
maxima $x_+<x_0<x_-$ there are two flucton solutions, also denoted by $\pm$,
which start at $x_0$ and ``relax" for an infinitely long time near either $x_+$ or $x_-$.
But now there is no symmetry $x\rightarrow -x$, these two fluctons have different energy
and the issue of relative normalization of their contributions is rather nontrivial.

The flucton path $f_+(\tau)$, starting from $x_+$, can reach any point we discuss. But,
an additional problem indicating troubles with such an approach, is that the flucton
path $f_-$ cannot reach {\it all} points $x_0$ in the interval $x_+ < x_0 < x_-$
since it has the energy $E_-$ insufficient to ``climb"
all the way to $x_+$. This path can only reach to the turning point and get reflected back.
A periodic path starting and ending at $x_-$ is known as ``bounce" solution.

A complexification of the paths opens many new options.
Let us start with the generalizations of the flucton path $f_+(\tau)$, starting from $x_+$
with the energy $E_+$. The initial velocity at the top is zero, but, as for a skier at the
mountain top, there is a freedom to slide in any direction.

Where one would like to go? Along the real axes no other point
has the  sufficient height of the inverted potential,  so the real flucton path $f_+$
can reach any $x_0$, inside or outside of the interval indicated. The
classical action $S_+=\int_{x_+}^{x_0} dx p(x)$ is a monotonous function since $p(x)>0$:
therefore the corresponding amplitude  $exp(-S_+(x_0))$
decreases monotonously in both directions, from $x_+$. This contribution
must be included into the density $P(x_0)$, but it cannot be the only one.

\begin{figure}[htbp]
\begin{center}
\includegraphics[width=6cm]{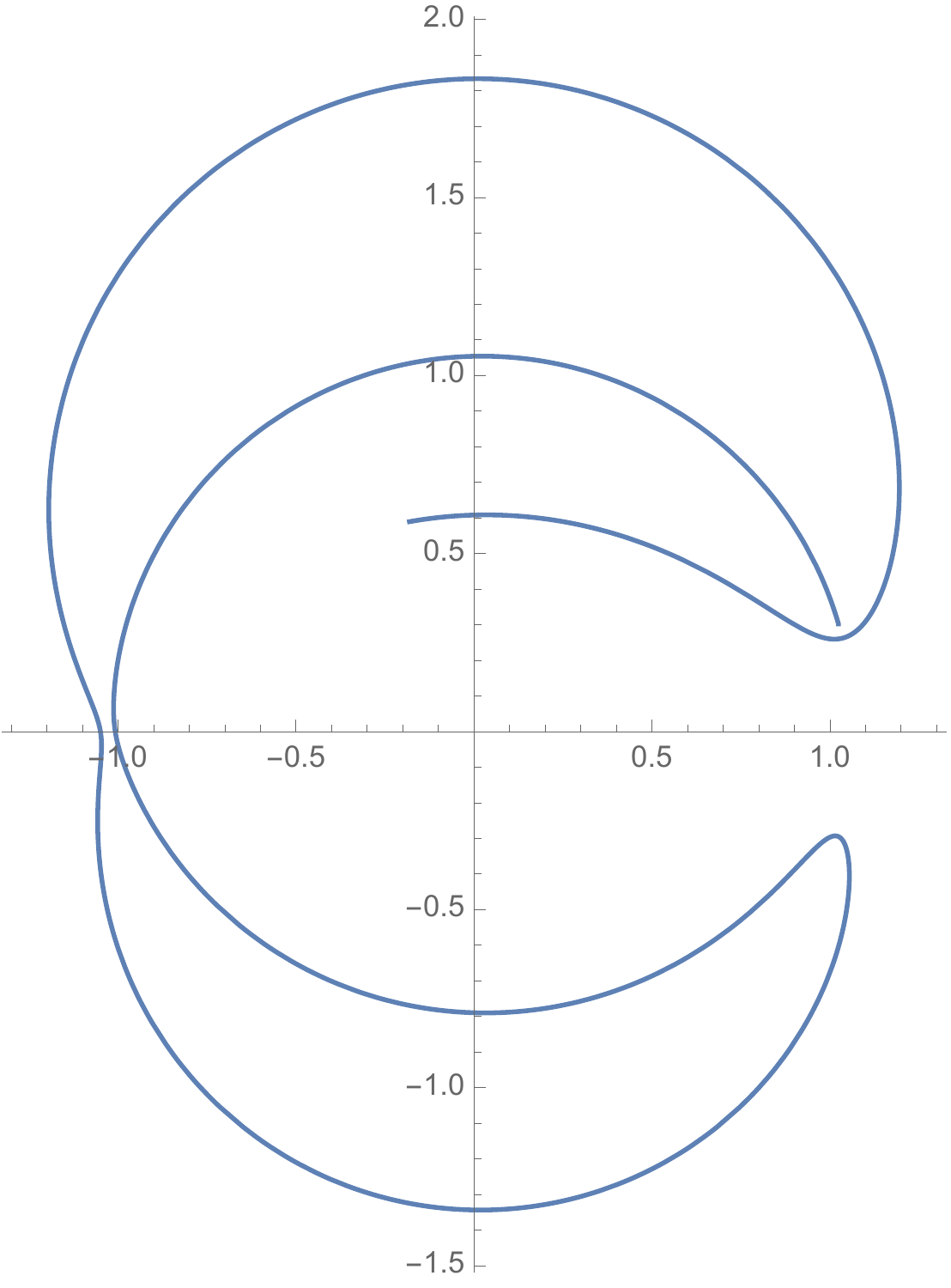}
\caption{Two examples of solutions to holomorphic
Newton's equation (\ref{hNE}). Both  start the zero velocity and slightly
displaced from the maximum of $-V$ location $x_+$.
The one going upward has a phase of the displacement tuned so that  it goes to the turning point	
$z_1$ and is reflected back on the same path, so that one see a single curve. This
 is the ``complex bion" of Ref.\cite{Kozcaz:2016wvy}.
The one going downward, from slightly different location, starts an infinite path
going around both turning paths $z_{1,2}$.
}
\label{fluct_plus_generic}
\end{center}
\end{figure}

Going into arbitrary direction from the global maximum leads to the family of paths, two of which is
shown in Fig. \ref{fluct_plus_generic}. They get reflected at (or a  vicinity) of two turning points  back to the maximum $x_+$. Since at its top ``relaxation" takes long time, such paths produce an option of being periodic, with an infinite period but finite action.

Let us find  the two turning points.
In the particular case of ADWP the potential is the 4th order polynomial
in coordinate, and thus it must have 4 roots. Hence, it can be
re-written in a form convenient for motion with the maximal energy $E_{+}=V(x_{+})$ as
\be
V-E_{+}\sim  (z-x_{+})^2(z-z_1)(z-z_2) \ .
\ee
Note that $x_{+}$ must be a double zero, and two others should be
the complex conjugated $z_1^*=z_2$ pair of two {\em turning points}.
(In our concrete example their location is at $z_{1,2}=1.02412 \pm 0.312482 i $.)
At these turning points the velocity on the path vanishes,
but since it is a not a maximum -- a double zero -- there is no long-time
``relaxation" possible, the ``complexified bion" path bounces back.


\begin{figure}[htbp]
\begin{center}
\includegraphics[width=8cm]{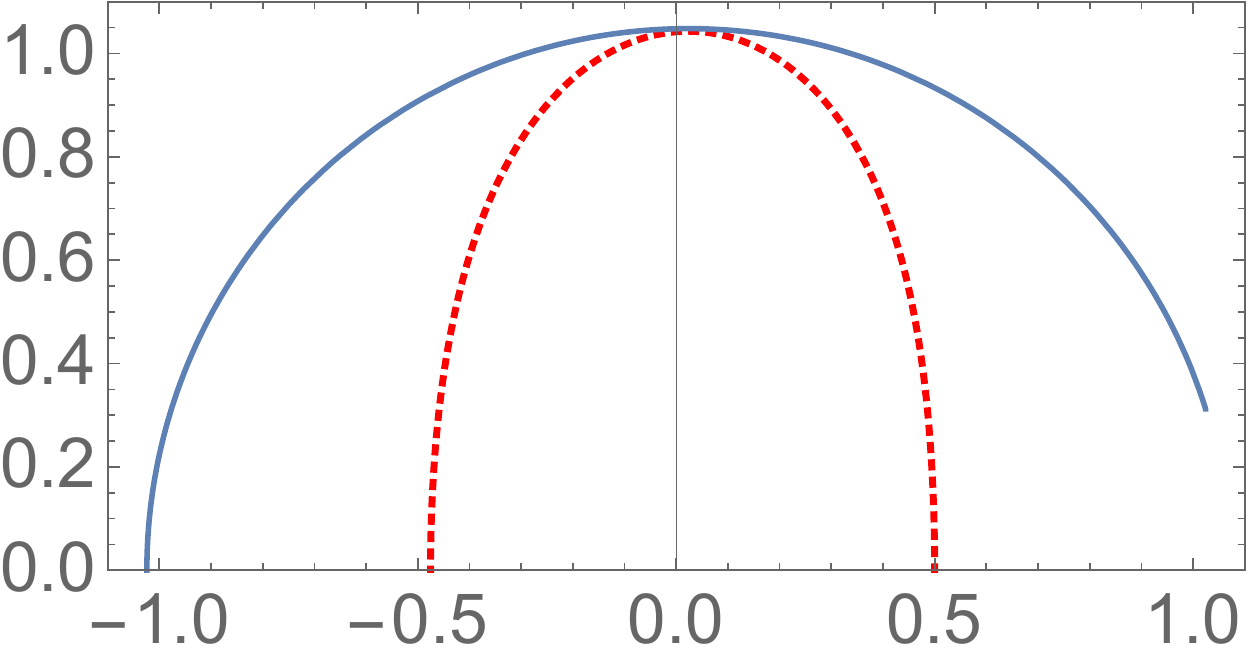}
\caption{Two solutions to the holomorphic
Newton's equation (\ref{hNE}). The
blue solid line is the same ``complex bion" of Ref.\cite{Kozcaz:2016wvy} as in the previous figure. The red dashed line starts at some generic point on the real axes, in this case $x_0=0.5$, with a velocity tuned so that it touches the
``complex bion".   }
\label{fig_fluct_complex}
\end{center}
\end{figure}

One of such periodic paths with finite action has been pointed out
in Refs. \cite{Kozcaz:2016wvy,Behtash:2015zha}
and named ``complex bion". The complex action of these paths contributing to the amplitude the factors $exp(-\mbox{Re} S(z_{CB})) cos( \mbox{Im} S(z_{CB}))$ produces cosine of certain non-trivial phases.

These phases violate the positivity of the amplitudes present for any real paths, and
produce interesting oscillations/cancellations.  Some known puzzles associated
with the energy spectra of quasi-exactly-solvable (QES) \cite{Turbiner:2016aum}
and/or supersymmetric (SUSY) examples  have been explained in these works.

Our aim is to find ``complex fluctons", classical paths connecting a generic point $x_0$
in the real axes with the global maximum at $x_+$. However, the paths belonging to the
family just described cross the real axes {\it only} at   $x_+$.

In general, starting with the real axes, one see that the kinetic energy $K=E-V(x_0)$ is real.
Therefore $\mbox{Im}(K)=\dot x \dot y=0$, which means that one of the factors must vanish. Thus there are two sets of paths: they
either go along the real axes, or normal to it.  The answer obviously is defined by  the sign
of the kinetic energy $2K=\dot x^2- \dot y^2$. Taking generic
initial point on the real axes as a starting point $x(t=0)=x_0$ and various initial values of $\dot y(t=0)$,
one can obtain  families of solutions to holomorphic EOM.

In Fig. \ref{fig_fluct_complex}
we show only one of them (red dotted line, for $x_0=0.5$) tuned to be touching the
``complex bion" path of Unsal et al (blue solid line). We propose to use a combination of
two segments of those two curves, before and after the touching point $z_{cross}$,
as a {\em ``complex flucton"}, leading from a generic point $x_0$ to the global maximum $x_+$. Note however, that while both curves at   the touching point $z_{cross}$ have the same
direction of the velocity, its magnitude needs to jump, as the two curves correspond to
two different energies.

For a generic paths one can think of them as classical ones, solving EOM
with appropriate external force term $z(\tau) f(\tau)$, added into the Lagrangian and to the EOM.
The advantage of the complex flucton path just introduced is that in this case the force should
only be applied at the crossing time moment only $f(\tau)\sim \delta(\tau-\tau_{cross})$,
as an instantaneous kick adjusting the total energy. There is a finite, although small
in the example considered, contribution of this kick to the action, which should not be omitted.

Summarizing a construction: there are two paths both leading from $x_0$ to the maximum $x_+$: the original real flucton and the complex one. The corresponding contributions to the density at $x_0$ are now both normalized in the same way, and their sum has the form
\be
    P(x_0)\sim 2 exp(-S_{real\, fl})+ \\
   exp[-\mbox{Re}(S_{complex \, fl})] 2 cos(\mbox{Im} (S_{complex \, fl}) \nonumber
\ee
where we have added the complex conjugate part of the path in the lower hemisphere  $\mbox{Im}(z)<0$.
Like for the complex bion contribution to the ground state energy, the contribution of the complex flucton may be positive or negative, depending on the particular value of the imaginary part of its action.

\section{Conclusions}

In our previous paper I we have outlined a new $flucton-based$ semiclassical theory,
based on path integral representation of the density matrix.  Corrections
to leading semiclassical results take the form of Feynman diagrams,  well defined
to any order by standard Feynman rules.
 As examples of its applications we
calculated one and two-loop corrections for the ground state density (square of the wave function) of the anharmonic oscillator (AHO).

In this second paper we describe the foundations of the method in more details, and also presented
a number of new results. At the start of the paper, we summarize the
completed one- and two-loop calculations based on Feynman diagrams, for all three physically
important examples: AHO, the double well (DWP) and sin-Gordon (SGP), see (\ref{eqn_results}).

We showed that in the case of polynomial potentials the perturbation corrections to the imaginary 
phase of wavefunction $\phi(x)$ are finite-degree polynomials. We demonstrated that generating functions 
of their leading degrees coincide to a corresponding terms in a loop expansion. Eventually, we found 
the Taylor expansion at small distances,
\[
  \phi(x)\ =\ A_0 x^2 + A_1 x^3 + \ldots
\]
while the loop expansion is noting but the expansion at large $x$. It is sufficiently straightforward 
to attempt to interpolate between these two regimes. For AHO the following interpolating trial function
\[
   \phi_{int}(x)\ =\ \frac{A + B x^2 + C g^2 x^4}{\sqrt {D + g^2 x^2}} + \frac{1}{2} 
   \log \text{Det}\,(O_{\text{flucton}})\ ,
\]
with parameters $A,B,C,D$
gives extremely high local accuracy for all $x$, in practice solving the problem, c.f. \cite{Turbiner:2010}. Similar approximants, interpolating between the small-$x$ series and
our loop expansion, can be done for other quantum-mechanical problems.

We also were able to relate these results to the iterative solution of certain equation (46)
for the reduced logarithmic derivative of the wave function. As we show explicitly, all our results
from Feynman diagrams  are reproduced exactly!  Among other insights, this way of deriving it
explains why all irrational functions -- such as logs and polylogs -- which appeared in expressions
for individual Feynman diagram, are always canceled in their sum. This way of calculation allows us
to go to higher orders: in particular, we calculated one more term of the expansion, corresponding
to the sum of the 15 three-loop Feynman diagrams, none of which was analytically evaluated so far.  
(Needless to say, this way of calculation, starting from  the Schr\"odinger  equation,
is not generalizable to QFT applications at least at present.)

We have studied the issues of convergence and the accuracy of this version of the semiclassical theory.
It is shown that in the case of weak coupling, the series for the density are well convergent.
Even if the coupling is not small, in the semiclassical domain it seems to be convergent
to the exact answer (provided one uses the logarithmic derivative
of the wave function, removing sensitivity to the normalization constant.)

Finally, we discussed the generalization of the flucton theory to the case of more than one
minimum of the potential using the example of the DWP problem. In the case when two minima are
degenerate, the density in between them is defined via a sum of 4 contributions:
left and right-side fluctons, the instanton and the anti-instanton solutions. Interestingly,
the latter produce constant ($x$-independent) contribution to the density matrix in this region.

The case of (slightly) non-degenerate minima is much more involved. The authors of
\cite{Kozcaz:2016wvy,Behtash:2015zha}, in which the ground state energy
of the asymmetric double well potential was studied, propose to complexify the coordinate
$x(\tau)\rightarrow z(\tau)=x(\tau)+i y(\tau)$ and include contributions of certain finite action
solution to the holomorphic EOM (\ref{hNE}). We studied such paths and found large families of such solutions, starting from the maximum of the inverted potential. We also show
how one can construct ``complexified fluctons", leading from a generic point to the global
maximum of the (inverted) potential, can be constructed using (segments of) $two$ such paths. While quantitative studies of those ``complex fluctons" are deferred to future works,
the main qualitative point is made here: since one of them has a complex
action,  its contribution to the density matrix has a nontrivial phase.
It can be either positive or negative, depending on the magnitude of the imaginary part of their action, depending in turn on the asymmetry parameter.

\begin{acknowledgments}

This work in its early stage was supported in part by CONACYT grant {\bf 166189}~(Mexico) for M.A.E.R. and A.V.T., and also by DGAPA grant {\bf IN108815}~(Mexico) for A.V.T. \quad
M.A.E.R. is grateful to ICN-UNAM, Mexico as well as the Stony Brook University for the kind hospitality during his visits, where a part of the research was done, he was supported in part by DGAPA grant
{\bf IN108815}~(Mexico) and, in general, by CONACyT grant {\bf 250881}~(Mexico) for postdoctoral research.
The work of E.S. is supported in part by the U.S. Department of Energy under Contract
No. {\bf DE-FG-88ER40388}.
The authors are grateful to participants of the seminar at Center for Nuclear Theory, Stony Brook University (Dec.21, 2016) for useful discussions.

\end{acknowledgments}

\end{document}